\documentclass[a4paper, 12pt]{article}
\usepackage{comment} 
\usepackage{lipsum} 
\usepackage{fullpage} 
\usepackage{booktabs}
\usepackage{setspace}
\usepackage{bbm}
\usepackage[hidelinks]{hyperref}
\usepackage{booktabs}
\usepackage[english]{babel}
\usepackage[utf8x]{inputenc}
\usepackage[margin=1in]{geometry}
\usepackage{graphicx}
\usepackage{subcaption}
\graphicspath{{desktop/}}
\usepackage[colorinlistoftodos]{todonotes}
\usepackage{amsmath}
\usepackage{amsthm}
\usepackage{amssymb}
\usepackage{mathrsfs}
\usepackage[colorinlistoftodos]{todonotes}
\usepackage[export]{adjustbox}
\usepackage{setspace}
\usepackage{natbib}
\setcitestyle{aysep={,}} 
\usepackage{pdflscape}
\usepackage{booktabs,caption}
\usepackage[flushleft]{threeparttable}
\usepackage{afterpage}
\makeatletter
\def\@seccntformat#1{\@ifundefined{#1@cntformat}%
   {\csname the#1\endcsname\space}
   {\csname #1@cntformat\endcsname}}
\newcommand\section@cntformat{\thesection.\space}       
\newcommand\subsection@cntformat{\thesubsection.\space} 
\makeatother

\usepackage[toc,page]{appendix}
\usepackage{scalerel,stackengine}
\stackMath
\newcommand\reallywidehat[1]{%
\savestack{\tmpbox}{\stretchto{%
  \scaleto{%
    \scalerel*[\widthof{\ensuremath{#1}}]{\kern.1pt\mathchar"0362\kern.1pt}%
    {\rule{0ex}{\textheight}}
  }{\textheight}%
}{2.4ex}}%
\stackon[-6.9pt]{#1}{\tmpbox}%
}
\renewenvironment{abstract}
 {\small
  \begin{center}
  \bfseries \abstractname\vspace{-.5em}\vspace{0pt}
  \end{center}
  \list{}{
    \setlength{\leftmargin}{.5cm}%
    \setlength{\rightmargin}{\leftmargin}%
  }%
  \item\relax}
 {\endlist}

\newtheorem{lem}{Lemma}
\usepackage{xpatch}
\makeatletter
\AtBeginDocument{\xpatchcmd{\@thm}{\thm@headpunct{.}}{\thm@headpunct{}}{}{}}
\makeatother

\makeatletter
\def\@seccntformat#1{\@ifundefined{#1@cntformat}%
   {\csname the#1\endcsname\quad}
   {\csname #1@cntformat\endcsname}
}
\makeatother

\title{
Identifying and Estimating Perceived Returns to Binary Investments%
\footnote{I thank Mary Kate Batistich, Trevor Gallen, Kendall Kennedy, Soojin Kim, Dan Millimet, Kevin Mumford, Victoria Prowse, and Miguel Sarzosa as well as seminar participants at Case Western Reserve University, The European Association of Labor Economists Meeting, Kansas State University, The Midwest Economics Association Meeting, The National Tax Association Meeting, Purdue University, The Southern Economic Association Meeting, and The US Census Bureau for helpful comments.}
}

\author{Clint Harris%
\footnote{Wisconsin Institute for Discovery, University of Wisconsin-Madison, 330 N Orchard Street,
Madison, WI 53715 USA; email: clint.harris@wisc.edu}}

\date{\today}

\begin{document}
\maketitle

\begin{abstract}
I describe a method for estimating agents' perceived returns to investments that relies on cross-sectional data containing binary choices and prices, where prices may be imperfectly known to agents.
This method identifies the scale of perceived returns by assuming agent knowledge of an identity that relates profits, revenues, and costs rather than by eliciting or assuming agent beliefs about structural parameters that are estimated by researchers. 
With this assumption, modest adjustments to standard binary choice estimators enable consistent estimation of perceived returns when using price instruments that are uncorrelated with unobserved determinants of agents' price misperceptions as well as other unobserved determinants of their perceived returns.
I demonstrate the method, and the importance of using price variation that is known to agents, in a series of data simulations.

\hfill

\end{abstract}

\clearpage





\section{Introduction}

In this paper I describe a method for estimating distributions of perceived private returns to binary investments. These structural perceived returns estimates are of distributions of agents' compensating variation associated with a binary choice that condition on observables. This method complements program evaluation methods that estimate effects of specific policy shocks on binary choices by allowing for predictions of counterfactual policies that differ from past policies in magnitude or targeted population.
For instance, \cite{h20} applies this method to estimate perceived returns to college, allowing for counterfactual predictions of targeted college attendance subsidies (and taxes) for diverse groups of individuals.
Identification is achieved by assuming common agent knowledge of an identity that relates prices to returns, while also using instruments that are de facto known to agents, in the sense that they shift perceived prices the same amount that they shift actual prices, in addition to satisfying the traditional exclusion restriction.

This paper presents a special case of a general method for identifying the scale of binary choice models by assuming agent beliefs about a variable observed by the researcher and agent beliefs about the mapping between that variable and the perceived return latent variable. Existing work that makes such assumptions includes \cite{Chn05}, who assume agent knowledge of their lifetime pecuniary return to college insofar as it is attributable to explanatory variables observed by the researcher, and \cite{dm18}, who assume partial agent knowledge of trade revenues and agent knowledge of an estimated demand elasticity parameter. The present paper assumes partial agent knowledge of prices in the sense of \cite{dm18} while assuming agent knowledge that prices causally decrease returns dollar for dollar in accordance with an identity that relates profits, revenues, and costs. The use of this identity imposes a theoretical restriction on a structural parameter (the coefficient on price in the binary choice latent variable equation) without requiring its estimation by researchers or agents. Avoiding the assumption that agents obtain the same estimate of a parameter as researchers improves robustness to the concerns articulated by \cite{manski1993adolescent, manski2004measuring} about the pitfalls of making incorrect assumptions on agents' knowledge of structural models.

The method in the present paper avoids assuming rational expectations on any model objects, instead assuming that the variation in prices associated with chosen instruments is known to agents regardless of whether agents are correct about prices on average. This makes it particularly attractive in applications where rational expectations assumptions in general are suspect, but the researcher can credibly argue that a particular price shock is nonetheless known to agents. Considering the example of college attendance, it is possible that exogeneous policy shocks may shift prices more than they shift perceived prices, as with Pell grants \citep{hansen1983impact, kane1995rising}, they may shift perceived prices more than they shift prices, as with the Michigan HAIL policy \citep{dlmo18}, or they may shift prices and perceived prices the same amount, as with the Social Security Student Benefit termination \citep{D03}. Of these preceding sources of variation, only the last would be appropriate for estimating the model presented in this paper. In addition to college attendance, attractive targets for this method include healthcare, home purchases, R\&D, and export decisions due to the substantial information frictions on prices in these settings.

In addition to considerations regarding the relative credibility of different assumptions on agent beliefs, applications also differ in data availability. The method described in this paper relies on cross-sectional data that contains binary choices on investments and prices associated with those investments. Methods that rely on rational expectations on ex post returns to investments require longitudinal data (without requiring data on prices), as in \cite{Chn05} and related research surveyed by \cite{Ch07}. Meanwhile, inferring beliefs by eliciting them directly from agents requires surveys that contain this information, as in \cite{J10}, \cite{Wz15}, and \cite{bz18}. The method described in this paper is thus useful in settings where there is no clear winner in terms of assumption validity, but when longitudinal data and data on agent perceptions in unavailable.


I describe how to estimate perceived returns when prices are known to agents and exogenous, and how to overcome violations of these conditions using instrumental variables. I compare performance of these methods with valid and invalid instruments across data generating processes that differ in the assumptions on agent knowledge of prices. In the most realistic settings, methods that make no use of instruments, or which use instruments that are correlated with agent misperceptions, perform poorly compared to those that use instruments that are de facto known to agents.

The plan of the rest of this paper is as follows. Section \ref{model} introduces the empirical model. Section \ref{empirical_strat} describes the econometric strategy and the assumptions required for identification. Section \ref{Simulations} evaluates the robustness of various methods and instruments to various empirical challenges in a series of simulated data exercises. Section \ref{Conclusion} concludes.

\section{Model}\label{model}
I assume that agents choose whether to make an investment based on their
beliefs about discounted net incomes and costs associated with choices, which I present as a two-sector generalized Roy \citeyearpar{r51} model.
Agents choose to select the investment, $S_i=1$, or to not do so, $S_i=0$, which is observed by the researcher. I define $\widetilde{Y}_{1,i}$ as agent $i$'s perceived discounted present value of lifetime income associated with choosing the investment and $\widetilde{Y}_{0,i}$ as their perceived discounted present value of lifetime income associated with not doing so. I further define $\widetilde{C}_i$ as their perceived net present value cost of making the investment, which includes prices paid and nonpecuniary costs expressed in monetary values. Unlike common applications of the Roy model, none of $\widetilde{Y}_{1,i}$, $\widetilde{Y}_{0,i}$, and $\widetilde{C}_i$ are observed by the researcher for any individual because they represent agent perceptions.


I express the perceived potential incomes and costs for individual $i$ with the following linear-in-parameters production functions,
\begin{equation}\label{potential_outcomes}
\begin{split}
\widetilde{Y}_{1,i} = &
X_i\beta_1+\tilde{\epsilon}_{1,i} \\
\widetilde{Y}_{0,i} = &
X_i\beta_0+\tilde{\epsilon}_{0,i} \\
\widetilde{C}_{i} = &
X_i\beta_C+\widetilde{Price}_i{}+\tilde{\epsilon}_{Ci}.
\end{split}
\end{equation}
Here, $X_i$ are variables observed by the researcher that determine potential incomes and costs. The parameters $\{\beta\}$ capture the extent to which these variables drive beliefs about potential outcomes regardless of whether they are known to agents. $\widetilde{Price}_i$ is the agent's perceived price for the investment, which is known to agents but not to researchers. Importantly, it is assumed to only affect costs and has a coefficient that is normalized to unity.
Finally, $\tilde{\epsilon}_{1,i}$, $\tilde{\epsilon}_{0,i}$, and $\tilde{\epsilon}_{Ci}$ represent idiosyncratic perceived returns to investment that are known to agents but not to the researcher.

I assume that agents maximize expected wealth independently of how they consume it, as in the case of perfect credit markets. It follows that the perceived net return/profit, $\widetilde{\pi}_i$,
is sufficient to determine agents' decisions in accordance with the rule \hfill
\begin{equation}\label{Selection_general}
S_i= 
\begin{cases} 
& 1 \mbox{, if } \widetilde{\pi}_i \geq 0, \\
& 0 \mbox{, otherwise.}
\end{cases}
\end{equation}
I further assume that the definition of profit, $\pi_i \equiv Revenue_i-Cost_i$, is known to agents in the sense that it holds for their beliefs as well, such that \hfill
\begin{equation}\label{profit_identity}
\begin{split}
\widetilde{\pi}_i &= \widetilde{Revenue}_i-\widetilde{Cost}_i \\
&= \widetilde{Y}_{1,i}-(\widetilde{Y}_{0,i}+\widetilde{C}_i),
\end{split}
\end{equation}
where $\widetilde{Revenue}_i$ denotes the agent's perceived income and $\widetilde{Cost}_i$ denotes the agent's perceived opportunity cost, which includes $\widetilde{Y}_{0,i}$.%
\footnote{I avoid denoting agents' beliefs with conditional expectations over realized values, as is common in the literature, to avoid the implication of rational expectations which follows from the law of iterated expectations.}
It follows that the agent's decision rule can be expressed in terms of potential outcomes as \hfill
\begin{equation}\label{Selection}
S_i= 
\begin{cases} 
& 1 \mbox{, if } \widetilde{Y}_{1,i}-\widetilde{Y}_{0,i}-\widetilde{C}_i \geq 0, \\
& 0 \mbox{, otherwise.}
\end{cases}
\end{equation}

Defining the net marginal effects $\beta \equiv \beta_1-\beta_0-\beta_C$ and the net idiosyncratic component of perceived outcomes $\tilde{\epsilon}_i \equiv \tilde{\epsilon}_{1,i}-\tilde{\epsilon}_{0,i}-\tilde{\epsilon}_{Ci}$, we can combine (\ref{potential_outcomes}) with (\ref{Selection}) to write the perceived return latent variable as \hfill
\begin{equation}\label{perceivedreturns}
\widetilde{\pi}_i = X_{i}\beta-\widetilde{Price}_i{}+\tilde{\epsilon}_{i}.
\end{equation}
Importantly, the assumptions given result in the latent variable being linear in perceived prices, with a marginal effect ($-1$) that is known to both agents and the researcher.%
\footnote{The researcher constraining the price coefficient to the value used by agents is key to identification, not the researcher or agents being correct about its value.}
The expression of perceived returns as a latent variable in a binary choice problem with a single known marginal effect is the starting point of the estimation procedures described below.

\section{Empirical Strategy}\label{empirical_strat}

It follows from the model that latent perceived returns are identified by $\beta$, $\widetilde{Price}_i$, and $\tilde{\epsilon}_i$, given the observed $X_i$. The lack of observation of $\tilde{\epsilon}_i$
is a common problem that will be addressed with commonly used binary choice estimation techniques. In this section I will describe adjustments to these estimators that leverage the assumptions described above 
to permit identification of $\beta$ and the scale of the distribution of $\tilde{\epsilon}_i$ in the context of the researcher's failure to observe agents' perceived prices. To preface, these adjustments address challenges that arise due to perceived costs having a causal effect on perceived returns in the identity given in (\ref{profit_identity}).

The econometric methods described below establish conditions under which the assumed coefficient on perceived prices from (\ref{perceivedreturns}) exactly determines the marginal effect of realized prices on perceived returns in a binary choice model. Omitted variable bias and measurement error in prices as measures of perceived prices threaten the validity of this assumption. It follows that methods which address omitted variable bias and measurement error will validate the assumption on the marginal effect of realized prices on perceived returns. To clarify, consider the expression of agents' beliefs about prices used throughout this paper,
\hfill
\begin{equation}\label{belief_restriction}
\widetilde{Price}_i = {Price}_i+X_i\alpha+\nu_i,
\end{equation}
where the realized price, $Price_i$, is observed by the researcher, $\alpha$ gives the effect of explanatory variables on price misperceptions, and $\nu_i$ is the idiosyncratic component of agent $i$'s misperception of prices. Here, realized prices are assumed to increase agents' beliefs about prices at a known marginal rate of unity insofar as they are known to agents.

This expression allows us to present an empirically tractable version of perceived returns,
\begin{equation}\label{perceivedreturns_empirical}
\begin{split}
\widetilde{\pi}_i &= X_{i}\beta-\widetilde{Price}_i{}+\tilde{\epsilon}_{i} \\
&= X_{i}\beta-Price_i{}-X_i\alpha{}-\nu_i{} +\tilde{\epsilon}_{i},
\end{split}
\end{equation}
by substituting in prices observed by the researcher for agents' unobserved perceived prices and defining $\theta=\beta-\alpha{}$.
\footnote{The distinction between the extent to which each control contributes to misperceptions in prices, $\alpha$, and to other components of perceived returns, $\beta$, is presented to emphasize that the methods in this paper are robust to systematic bias in perceptions associated with explanatory variables, even though they are not separately identified.} This representation presents the unexplained price misperception as an omitted variable, which will produce problems if $Price_i$ is correlated with $\nu_i$. Natural examples of problematic correlations between price misperceptions include agents systematically over-reacting or under-reacting to price predictors that are unobserved by the researcher. The extreme case of under-reaction is that in which an unobserved predictor of realized price variation is ignored by or unknown to agents altogether, which amounts to classical measurement error in realized prices as measures of perceived prices.

In what follows, I first consider a benchmark case in which unobserved components of price misperceptions are mean independent of realized prices and prices are uncorrelated with unobserved determinants of perceived returns. Though agents may be mistaken about prices, actual prices can stand in for perceived prices because any systematic price misperceptions are accounted for by observables. Second, I consider the case in which prices are correlated with unobserved price misperceptions and unobserved components of perceived returns. In this setting, instruments for observed prices that are uncorrelated with unobserved components of perceived returns will be needed to identify perceived returns. This case emphasizes the importance of choosing instruments that are de facto known to agents in addition to being exogenous for constructing credible counterfactuals relating to price changes.

\subsection{Estimation with Known, Exogenous Prices}\label{MLE_method}
Here, I describe a benchmark procedure for estimating perceived returns with a simple adjustment to a common binary choice method. This procedure will provide consistent estimates of the perceived returns distribution under two assumptions that are likely to be violated in applications. First, this method assumes that prices and the unobserved component of perceived returns are uncorrelated. Second, it assumes that unobserved components of price misperceptions are mean independent of prices conditional on $X_i$, the simplest case of which is agents having perfect information on prices.

With the decision rule in (\ref{Selection}) and the expression of perceived returns in (\ref{perceivedreturns_empirical}), an assumption on the distribution of $-\nu_i{} +\tilde{\epsilon}_{i}$ is sufficient to consistently estimate perceived returns by maximum likelihood. I assume the composite unobserved component of perceived returns in (\ref{perceivedreturns_empirical}) is normally distributed as
\begin{equation}
    -\nu_i{} +\tilde{\epsilon}_{i}|X_i,Price_i \sim \mathcal{N}(0,\sigma^2).
\end{equation}
The assumption of normality is chosen for convenience, and is not necessary for the estimation procedures in this paper. Defining $(\beta^*,\theta^*,{\gamma}^*)=(\frac{\beta}{\sigma},\frac{\theta}{\sigma}, \frac{1}{\sigma})$ for notational convenience, the probability of selection is given by
\begin{equation}
Pr(S_i=1|X_i,Price_i) = \Phi
(
X_i\theta^*-Price_i{\gamma}^*
)
,
\end{equation}
where $\Phi(\cdot)$ denotes the standard normal CDF. 

The parameters $(\theta^*,{\gamma}^*)$ are the values that maximize the log-likelihood
\begin{equation}\label{L-likelihood}
\begin{gathered}
\mathcal{L}(\theta^*,{\gamma}^* |
X_i,Price_i) = \\
\sum_i 
S_i\log\Bigg[
\Phi
\Big
(
X_i\theta^*-Price_i{\gamma}^*
\Big
)
\Bigg]
+ (1-S_i)\log\Bigg[
1-\Phi\Big(
X_i\theta^*-Price_i{\gamma}^*
\Big)\Bigg].
\end{gathered}
\end{equation}
The estimates of perceived returns are then given by
\begin{equation}\label{probitdist}
\hat{\widetilde{\pi}}_i|X_i, Price_i \sim \mathcal{N}(X_{i}\hat{\theta}-Price_i{},\hat{\sigma}^2),
\end{equation}
where imposing the constraint $\gamma^*=\frac{1}{\sigma}$ (rather than the standard constraint $\sigma=1$) is the only difference from a standard probit. Importantly, the assumption that $\gamma^*=\frac{1}{\sigma}$ is only valid under the assumptions described in Section \ref{model} when realized prices are uncorrelated with unobserved components of price misperceptions and perceived returns conditional on $X_i$. As this generally will not be the case, this assumption is not an innocuous normalization.

\subsection{Estimation with Endogenous, Unknown Prices}\label{CF_method}
Here, I describe a control function approach that addresses correlation between prices and unobserved components of perceived returns as well as arbitrary correlation between prices and misperceptions on prices. In Appendix \ref{appendix:Moment_Inequalities}, I discuss a method developed by \cite{dm18} that performs well in this model when agents under-react to price variation, such as when they form rational expectations on prices based on a known price predictors and only a subset of price predictors are known to them. The method in this section uses an established estimator, but adds the assumption that instruments are uncorrelated with unobserved components of price misperceptions in addition to the more commonly invoked assumption that instruments are uncorrelated with other unobserved idiosyncratic components of perceived returns. This additional assumption contributes to credibility for predictions of responds to counterfactual price changes that are known to agents, without changing the asymptotic or finite sample properties of the estimator. 

The control function approach uses the following system of equations, with reference to the expression of perceived returns in (\ref{perceivedreturns_empirical}),\hfill
\begin{equation}\label{Price_Z}
\begin{gathered}
\widetilde{\pi}_i = X_i\theta-{Price_i}{}-\nu_i{}+\tilde{\epsilon}_i \\
Price_i =
Z_i\delta +u_i,
\end{gathered}
\end{equation}
where I have left unobserved price misperceptions and other unobserved components of perceived returns separate for clarity. Here, I introduce the instruments, $Z_i$, where $X_i \subset Z_i$, that are assumed to be conditionally uncorrelated with $-\nu_i+\tilde{\epsilon}_i$ and strongly correlated with observed prices. With some loss of generality, I will refer to instruments that satisfy this condition as ``known and exogoneous'' for brevity.%
\footnote{It is not necessary that agents know the instruments in $Z_i$, but only that they know the variation in prices that is attributable to $Z_i$. For example, agents need not know about a tax or subsidy shock to the price of investment, so long as they are aware of the change in price that arises from the policy shock. Furthermore, the language that instruments are known and exogenous suggests that $Cov(Z_i,\nu_i)=Cov(Z_i,\tilde{\epsilon}_i)=0$, while these are sufficient but not necessary for the less intuitive condition $Cov(Z_i,\tilde{\epsilon}_i-\nu_i)=0$, which accommodates the knife-edge case of the two sources of bias cancelling out.}
With valid instruments, the price residual $u_i$ contains all components of prices that are correlated with idiosyncratic components of price misperceptions or other unobserved components of perceived returns.

Given the above, I estimate the following equation,
\begin{equation}\label{Perceived_Returns_CF}
\begin{split}
\widetilde{\pi}_i 
=& X_i\theta-{Price_i}{}-\nu_i{}+\tilde{\epsilon}_i \\
=& X_i\theta-{Price_i}{}+u_i\rho+\xi_i \\
=& X_i\theta-{Price_i}{}+\hat u_i\rho+\zeta_i.
\end{split}
\end{equation}
The first line follows directly from the representation of perceived returns in (\ref{perceivedreturns_empirical}). 
The second line substitutes in the linear projection of the composite error $-\nu_i{}+\tilde{\epsilon}_i$ on the first stage error $u_i$, wherein $\rho = \mathbb{E}[u_i(-\nu_i{}+\tilde{\epsilon}_i)]/\mathbb{E}[u_i^2]$ and $\xi_i$ is the residual when controlling for $u_i$. The third line substitutes the estimated residuals from the first stage regression of $Price_i$ on $Z_i$ in for their unobserved true values, generating a new error, $\zeta_i = \xi_i+(u_i-\hat u_i)\rho$. This new error will converge asymptotically to $\xi_i$, but will differ in small samples due to sampling error in the estimation of the residual from the first stage, $\hat u_i$.

To estimate perceived returns, I assume that the new error in the perceived returns control function expression is normally distributed,
\begin{equation}\label{error_CF}
    \zeta_i|X_i,Price_i,\hat u_i \sim \mathcal{N}(0,\sigma_\zeta^2),
\end{equation}
noting that the variance of $\zeta_i$ will differ from that of $\tilde{\epsilon}_i$ if $\rho\neq0$.
I estimate perceived returns using two-stage conditional maximum likelihood, following \cite{Rv88}, while correcting for the inclusion of estimated regressors, following \cite{mt85}, though other estimators will also provide consistent estimates.
Defining $(\theta^*_\zeta,{\gamma}^*_\zeta,\rho^*_\zeta)=(\frac{\theta}{\sigma_\zeta},\frac{{1}}{\sigma_\zeta}, \frac{\rho}{\sigma_\zeta})$, the log-likelihood for the second stage of the control function approach is given by%
\footnote{As an closely-related alternative, we could perform a instrumental variables probit to obtain identical estimates of $\theta$. The control function method has the advantage of conditioning on the variation in prices that isn't used in identifying the effect on perceived returns, which permits more precise counterfactual predictions for policies that are targeted on observables.
}
\begin{equation}
\begin{gathered}
\mathcal{L}\Big(\theta^*,{\gamma}^*,\rho^* |
X_i, \hat u_i \Big) = \\
\sum_i 
S_i\log\Bigg[\Phi\Big(X_i\theta^*_\zeta-{Price_i}{\gamma}^*_\zeta+\hat u_i \rho^*_\zeta
\Big)
\Bigg] \\
+ (1-S_i)\log\Bigg[1-\Phi\Big(X_i\theta^*_\zeta-{Price_i}{\gamma}^*_\zeta+\hat u_i\rho^*_\zeta
\Big
)\Bigg].
\end{gathered}
\end{equation}
Estimates of perceived returns are obtained by plugging the estimated parameters and the assumed coefficient on perceived prices into the latent variable equation,
\begin{equation}\label{CF_dist}
\widetilde{\pi}_i|X_i,\hat u_i \sim \mathcal{N}\Big(X_i\hat{\theta}-{Price_i}{}+\hat u_i\hat{\rho},\hat{\sigma}_{\zeta}^2\Big).
\end{equation}

\section{Simulations}\label{Simulations}
In this section I apply the methods described above to simulated datasets to compare their performance. The important considerations involve agent beliefs about prices, price endogeneity, and instruments being known and/or exogenous to agents. Because the estimators used are standard, I stop short of performing full Monte Carlo simulations, instead comparing the performance instruments according to whether they are known or exogenous to agents within individual simulations. For additional simulations which compare the methods of this paper to the method of \cite{dm18}, see Appendix \ref{appendix:More_Sims}.

For the simulations, I use the following DGP, 
\hfill
\begin{equation}
\begin{gathered}
\widetilde{\pi}_i = X_i\beta-\widetilde{Price_i}{}+\tilde{\epsilon}_i \\
\widetilde{Price}_i = {Price}_i+\nu_i=Z_i\delta +u_i +\nu_i, \\
\end{gathered}
\end{equation}
where the nature of the covariance of $(Z_i,u_i,\nu_i,\tilde{\epsilon}_i)$ will determine the performance of various estimation approaches.
Both the probit and the control function method will obtain estimates of $\beta$, while the probit will estimate
\begin{equation}
\sigma = Var(-\nu_i+\epsilon)
\end{equation}
and the control function method will estimate
\begin{equation}
\begin{gathered}
\rho = \mathbb{E}[u_i(-\nu_i{}+\tilde{\epsilon}_i)]/\mathbb{E}[u_i^2],\\
\sigma_\zeta = \sqrt{Var(\zeta_i)} = \sqrt{Var(-\nu_i{}+\tilde{\epsilon}_i-\hat u_i\rho)}.
\end{gathered}
\end{equation}
Each DGP is comprised of $N=10,000$ observations of agents whose decisions are governed by their perceived returns to investment.

\subsection{Simulation with Known, Exogenous Prices}
I begin with a well-behaved benchmark DGP that corresponds to the setting described in Section \ref{MLE_method}.
I generate data according to \hfill
\begin{equation}
\begin{bmatrix}
z_i \\ 
u_i \\
\nu_i \\
\tilde{\epsilon}_i
\end{bmatrix}
\sim 
\mathcal{N}(\textbf{0},\Sigma);
\quad \Sigma = 
\begin{bmatrix}
 & 4 & 0 & 0 & 0 &\\
& 0 & 1 & 0 & 0 &\\
 & 0 & 0 & 2 & 0 & \\
 & 0& 0& 0 & 2 &
\end{bmatrix}.
\end{equation}
I construct the instrument vector as $Z_i = [X_i \mbox{ } z_{1,i}]$ where $X_i$ includes only a constant, and $\alpha=0$ such that $\theta=\beta$. Finally, I set $\beta=1$ and $\delta=[0 \mbox{ } 1]'$. Although I set $Var(\nu_i) = 2$, I describe prices as known in this setting because the price misperception is uncorrelated with prices.\footnote{This setting is one in which agents are wrong about prices in ways that are unrelated to price determinants. This sort of price misperception is plausible in cases where prices change frequently according to a distribution that is de facto known to agents, such as frequently repeated investments.}

\begin{figure}[hbtp!]
\centering
\includegraphics[width=.8\linewidth]{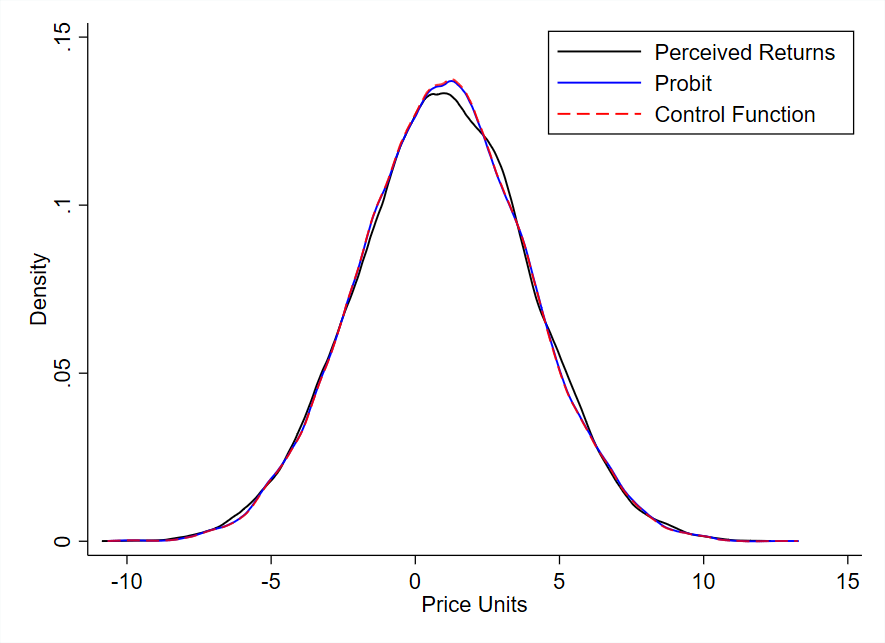}
\begin{minipage}{1\linewidth}
\caption{Simulation 1, Implied Perceived Returns Distributions}
\label{fig:sim_main_1}
\smallskip
\footnotesize
\emph{Notes:} Estimated densities of perceived returns given by the probit method using expression (\ref{probitdist}), and the control function method using expression (\ref{CF_dist}).
\end{minipage}
\end{figure}

Table \ref{tab:sim_main_1} shows perceived returns estimates for one simulation of this DGP using the methods from Section \ref{MLE_method} and Section \ref{CF_method}. Figure \ref{fig:sim_main_1} shows the distributions implied by the estimates for each method. In this case, the lack of correlation between prices and unobserved components of perceived returns, including price misperceptions, means that both methods will provide consistent estimates of perceived returns.

\begin{table}[htbp]\centering
\def\sym#1{\ifmmode^{#1}\else\(^{#1}\)\fi}
\caption{Simulation 1, Perceived Returns Estimates}
\label{tab:sim_main_1}
\begin{tabular}{l*{1}|c|*{3}{c}}
\hline\hline
                    &            &\multicolumn{1}{c}{(1)}&\multicolumn{1}{c}{(2)}\\
                    &   Target   &\multicolumn{1}{c}{Probit}&\multicolumn{1}{c}{Control Function}\\
\hline
Constant            &     1      &      0.959 &       0.955\\
                    &            &    (0.031) &     (0.031)\\
$\sigma$            &     2      &      1.922 &           .\\
                    &            &    (0.037) &            \\
$\sigma_\zeta$      &   1.977    &          . &       1.913\\
                    &            &            &     (0.039)\\
$\rho$              &     0      &          . &       0.021\\
                    &            &            &     (0.033)\\
\hline
Observations        &            &       10000&       10000\\

\hline\hline
\end{tabular}
\begin{minipage}{1\linewidth}
\smallskip
\footnotesize
\emph{Notes:} Standard errors in parentheses, corrected for the inclusion of estimated regressors following \cite{mt85} in the case of the control function. 
Parameters are in monetary units.
Estimates relate to expressions (\ref{probitdist}) and (\ref{CF_dist}), respectively.
All data is generated in Stata using random seed 1234.
\end{minipage}
\end{table}

\subsection{Simulation with Unknown, Endogenous Prices}
In this simulation, I consider a DGP that corresponds to the setting described in Section \ref{CF_method} in which agents systematically misperceive prices in ways that not accounted for by observables, and prices are correlated with unobserved components of perceived returns. I also compare the performance of an instrument that is exogenous but unknown to one that is both known and exogenous.
I generate data according to
\begin{equation}
\begin{bmatrix}
z_{1,i} \\ 
z_{2,i} \\
u_i \\
\nu_i \\
\tilde{\epsilon}_i 
\end{bmatrix}
\sim 
\mathcal{N}(\textbf{0},\Sigma);
\quad \Sigma = 
\begin{bmatrix}
&9 & 0 & 0 & -4& 0&\\
&0 & 9 & 0 & 0& 0&\\
&0 & 0 & 27 & -5& 9&\\
& -4 & 0 & -5 & 9& 0 & \\
& 0& 0& 9 & 0 & 16 &
\end{bmatrix}.
\end{equation}
I construct the instrument vector as $Z_i = [X_i \mbox{ } z_{1,i} \mbox{ } z_{2,i}]$ where $X_i$ includes only a constant, and $\alpha=0$ such that $\theta=\beta$. Finally, I set $\beta=1$ and $\delta=[0 \mbox{ } 1 \mbox{ } 1]'$.

In this case, there is positive correlation between $u_i$ and $\tilde{\epsilon}_i$ such that individuals who face idiosyncratically high prices also have high perceived returns, as may occur with price discrimination. Additionally, there is negative correlation between $u_i$ and $\nu_i$ such that individuals systematically underestimate the extent to which their price deviates from the average, as may occur if agents form rational expectations on prices conditional on an incomplete set of price determinants. Finally, this DGP includes two potential instruments; $z_{1,i}$, which is exogenous but not fully known to agents, as in the case of a poorly publicized policy shock, and $z_{2,i}$, which is both exogenous and known to agents.

Because $z_{1,i}$ is correlated with $\nu_i$, it is not a valid instrument for the purposes of this paper. For the control function estimates of $\rho$ and $\sigma_{\zeta}$, I use $u_{1,i}$ in place of $u_i$, where $u_{1,i} = z_{1,i}\delta_1+u_i$. In applications with many valid instruments, including different combinations of instruments will result in different estimates $\hat u_i$, $\hat \rho$ and $\hat \sigma_\zeta$, while nonetheless all returning consistent estimates of perceived returns. For comparisons between instruments, the complete distribution of perceived returns (succinctly described by the figures) and the estimated coefficients on $X_i$ will be correct for all valid instruments.

\begin{figure}[hbtp!]
\centering
\includegraphics[width=.8\linewidth]{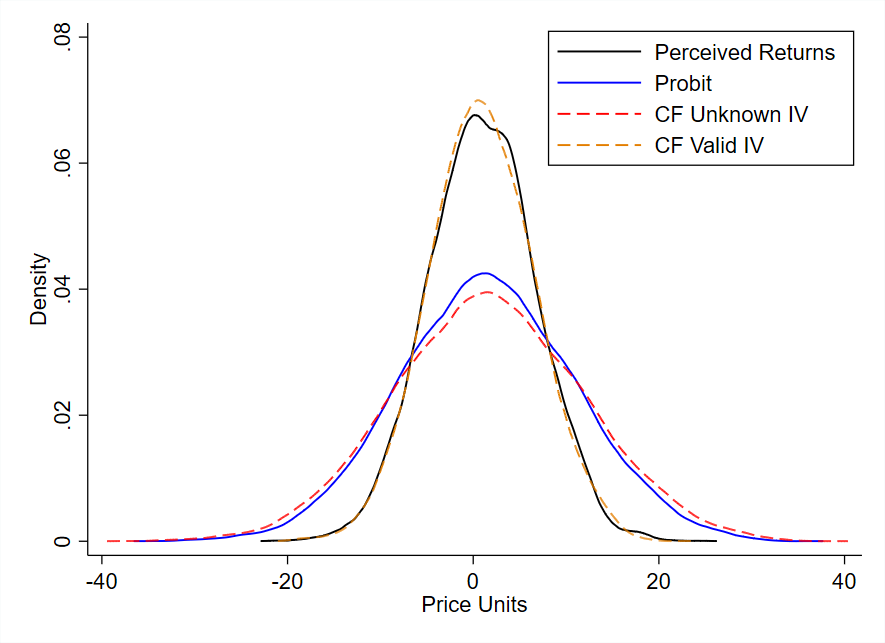}
\begin{minipage}{1\linewidth}
\caption{Simulation 2, Implied Perceived Returns Distributions}
\label{fig:sim_main_2}
\smallskip
\footnotesize
\emph{Notes:} Estimated densities of perceived returns given by the probit method using expression (\ref{probitdist}), and the control function method using expression (\ref{CF_dist}). The unknown IV is $z_{1,i}$ and the valid IV is $z_{2,i}$, where each IV is excluded from the estimation model when the other is used.
\end{minipage}
\end{figure}

Table \ref{tab:sim_main_2} shows the estimates for one simulation of this DGP using both methods, and also using each instrumental variable individually.  Figure \ref{fig:sim_main_2} shows the distributions implied by the estimates for each method. Because $z_{1,i}$ is correlated with misperceptions, it is not a valid instrument, and results in an estimated perceived returns distribution that is no better than that obtained when using no instruments.\footnote{For estimating instrument-specific intent to treat effects of prices on investment, which would be sufficient for determining the performance of a particular policy in the context of its actual implementation, instruments such as $z_{1,i}$ are valid. They nonetheless fail to provide credible insight into counterfactual policy changes that are well-publicized.}

\begin{table}[htbp]\centering
\def\sym#1{\ifmmode^{#1}\else\(^{#1}\)\fi}
\caption{Simulation 2, Perceived Returns Estimates}
\label{tab:sim_main_2}
\begin{tabular}{l*{1}|c|*{3}{c}}
\hline\hline
                    &            &\multicolumn{1}{c}{(1)}&\multicolumn{1}{c}{(2)}&\multicolumn{1}{c}{(3)}\\
                    &   Target   &\multicolumn{1}{c}{Probit}&\multicolumn{1}{c}{Control Function $z_1$}&\multicolumn{1}{c}{Control Function $z_2$}\\
\hline
Constant            &     1      &     1.528  &      1.659 &       0.863\\
                    &            &   (0.102)  &    (0.127) &     (0.086)\\
$\sigma$            &     5      &     7.001  &          . &           .\\
                    &            &   (0.138)  &            &            \\
$\sigma_\zeta$      &   3.954    &         .  &      7.552 &       3.927\\
                    &            &            &    (0.304) &     (0.126)\\
$\rho$              &     .5     &         .  &     -0.097 &       0.507\\
                    &            &            &    (0.046) &     (0.018)\\
\hline
Observations        &            &       10000&       10000&       10000\\

\hline\hline
\end{tabular}
\begin{minipage}{1\linewidth}
\smallskip
\footnotesize
\emph{Notes:} Standard errors in parentheses, corrected for the inclusion of estimated regressors following \cite{mt85} in the case of the control function. 
Parameters are in monetary units.
Estimates relate to expressions (\ref{probitdist}) and (\ref{CF_dist}), respectively.
All data is generated in Stata using random seed 1234.
\end{minipage}
\end{table}

\section{Conclusions}\label{Conclusion}
In this paper I describe how to estimate perceived returns to investments by assuming agent knowledge of an intuitive identity and modestly altering common estimation techniques. The assumption on agent knowledge may be preferable to rational expectations or related assumptions in applications. I further describe the econometric challenges that arise from the assumption and how to overcome them with careful choice of instruments that are not only exogenous to agents, but are also de facto known to them.

This method is relevant in many empirical questions, especially those subject to substantial information frictions on prices such as such as college attendance, firm R\&D, automobile purchases, home purchases, and healthcare. While the estimation techniques used in this paper are restricted to a probit and a control function probit, the general insights are relevant to more sophisticated models that involve responses to prices. Implementation of the identity relating perceived returns and prices used in this paper in the context of more sophisticated models, such as \cite{blp95} and its extensions, are left to future work.

In terms of policy implications, the methods described in this paper are relevant for constructing credible counterfactuals for well-publicized price changes, which are relevant for taxes and subsidies on investments including those associated with education and healthcare. The general insight is to avoid being too quick to assume that agents have rational expectations on model objects when alternative assumptions may be more defensible. Relatedly, the insights here also caution against extrapolating effects of counterfactual policies when the policy effects are estimated using a source of variation in prices that may not be known to agents. In practice, applied researchers should justify that sources of variation used for estimating treatment effects are known to agents just as they justify that they are exogenous to agents when making counterfactual predictions.

\clearpage 
\bibliographystyle{econometrica.bst}
\renewcommand{\bibname}{BIBLIOGRAPHY}
\bibliography{References}

\clearpage


\appendixpageoff
\begin{appendices}

\numberwithin{equation}{section}
\makeatletter 
\renewcommand{\section@cntformat}{Appendix \thesection:\ }
\makeatother

\renewcommand{\thesubsection}{\Alph{section}.\arabic{subsection}}

\setcounter{table}{0}
\renewcommand{\thetable}{\Alph{section}.\arabic{table}}

\setcounter{figure}{0}
\renewcommand{\thefigure}{\Alph{section}.\arabic{figure}}

\section{Moment Inequalities}\label{appendix:Moment_Inequalities}\label{MI_method}

This section describes how to adapt the moment inequality method developed by \cite{dm18} (DM) to the setting described in this paper. The setting of DM involves trade revenues that are partially observed by firms, which have a structural relationship with profits that is assumed to be known to agents. The type of information frictions described in DM are a special case of those described in the present paper, in which some sources of variation in the treatment variable are unknown to agents. In the context of the model presenting in equation (\ref{belief_restriction}), this involves negative correlation between $\nu_i$ and $Price_i$ such that prices are a mean preserving spread of perceived prices. Furthermore, the DM method assumes that ${\tilde{\epsilon}}_i$ is independent of other determinants of perceived returns.

This method makes use of instruments, $Z_i$ that are independent of $({\tilde{\epsilon}}_i, \nu_i)$. For $X_i \subset Z_i$, this implies that the expectation of (\ref{belief_restriction}) conditional on $Z_i$ gives \hfill
\begin{equation}\label{price_equal_perceived_Z}
 \mathbb{E}[\widetilde{Price}_i|Z_i]
 =
    \mathbb{E}[{Price}_i|Z_i]+X_i\alpha.
\end{equation}
Additionally, it makes a distributional assumption on unobserved perceived returns such as\hfill
\begin{equation}
    {\tilde{\epsilon}}_i | X_i, \widetilde{Price}_i \sim \mathcal{N}(0,\sigma_{\tilde{\epsilon}}^2),
\end{equation}
where the assumption of normality is unnecessary, but there are some restrictions on the assumed distribution which I discuss below. The method uses two types of moment inequalities to obtain bounds on the parameters of perceived returns, $(\theta,\sigma_{\tilde{\epsilon}})$. I will present the inequalities and provide a brief discussion here. For the derivation and further discussion of the moment inequalities, see DM. 

\subsection{Revealed Preference Moment Inequalities}

Defining $(\beta^*_{{\tilde{\epsilon}}},\theta^*_{{\tilde{\epsilon}}},{\gamma}^*_{{\tilde{\epsilon}}})=
(\frac{\beta}
{\sigma_{\tilde{\epsilon}}
},
\frac{\theta}{\sigma_{\tilde{\epsilon}}}
, \frac{{1}}{\sigma_{\tilde{\epsilon}}})
$ for notational convenience, the conditional revealed preference moment inequalities are
\begin{equation}\label{RPMI}
\begin{gathered}
\mathbb{E}\Bigg[
S_i (X_i\theta^*_{{\tilde{\epsilon}}}-Price_i{\gamma}^*_{{\tilde{\epsilon}}})+
(1-S_i)
\frac
{\phi(X_i\theta^*_{{\tilde{\epsilon}}}-Price_i{\gamma}^*_{{\tilde{\epsilon}}})}
{1-\Phi(X_i\theta^*_{{\tilde{\epsilon}}}-Price_i{\gamma}^*_{{\tilde{\epsilon}}})}
\Bigg|Z_i\Bigg]
\geq 0,\\
\mathbb{E}\Bigg[
-(1-S_i)(X_i\theta^*_{{\tilde{\epsilon}}}-Price_i{\gamma}^*_{{\tilde{\epsilon}}})+
S_i
\frac
{\phi(X_i\theta^*_{{\tilde{\epsilon}}}-Price_i{\gamma}^*_{{\tilde{\epsilon}}})}
{\Phi(X_i\theta^*_{{\tilde{\epsilon}}}-Price_i{\gamma}^*_{{\tilde{\epsilon}}})}
\Bigg|Z_i\Bigg]
\geq 0.
\end{gathered}
\end{equation}
These inequalities are consistent with the revealed preference argument that perceived returns are positive for those who select the investment and negative for those who do not.
Here, I provide an overview of the intuition.

Regarding the first inequality, consider an agent that selects the investment such that $S_i=1$. Following the revealed preference argument articulated in (\ref{Selection}) and the representation of perceived returns in (\ref{perceivedreturns_empirical}), it follows that this individual's perceived return is positive, such that \hfill 
\begin{equation}
S_i(X_{i}\theta-{Price}_i{}-\nu_i{}+{\tilde{\epsilon}}_{i})\geq 0.
\end{equation}
This expression cannot be computed directly because researchers do not observe $\nu_i$ or ${\tilde{\epsilon}}_i$.
However, as the inequality holds for all $i$, it follows that it holds in expectation conditional on $Z_i$,
\hfill
\begin{equation}
\mathbb{E}[
S_i
(
X_{i}\theta-{Price}_i{}
-
\nu_i{}
+
{\tilde{\epsilon}}_i
)
|
Z_i
]
\geq 0.
\end{equation}
Finally, it follows from the Law of Iterated Expectations, the assumption that $\nu_i$ is unknown to agents and therefore not acted upon, and the assumption that $Z_i$ is uncorrelated with $\nu_i$ such that $\mathbb{E}[S_i \nu_i {} | Z_i] = {}\mathbb{E}[S_i \mathbb{E}[\nu_i| S_i, Z_i]|Z_i] = 0$, yielding,\hfill%
\begin{equation}
\mathbb{E}[
S_i
(
X_{i}\theta-{Price}_i{}
+
{\tilde{\epsilon}}_i
)
|
Z_i
]
\geq 0.
\end{equation}
The first inequality in (\ref{RPMI}) is derived from this inequality where its second term is a positively biased approximation of $\mathbb{E}[S_i{\tilde{\epsilon}}_i|Z_i]$ that exploits the closed form for $\mathbb{E}[S_i{\tilde{\epsilon}}_i|X_i, \widetilde{Price}_i]$ under the normality assumption on ${\tilde{\epsilon}}_i$.\footnote{The bias makes substitution of prices for perceived prices nontrivial, and contributes to the inequality.}

Heuristically, if observed prices are a mean-preserving spread of perceived prices, substituting them in place of perceived prices will mistakenly increase expected perceived returns unconditional on selection for some agents and decrease them for others%
. 
For the agents for whom this expectation increases, the expectation of the error conditional on selection approaches zero%
. For those for whom it decreases, the expectation of the error conditional on selection approaches positive infinity%
. In many cases, this second effect will dominate the overall expectation of the error conditional on selection.\footnote{Global convexity of $\mathbb{E}[{\tilde{\epsilon}}_i|{\tilde{\epsilon}}_i<\kappa]$ in $\kappa$ is necessary for the inequalities to hold regardless of the value of $\kappa$ and the variance of the misperception term. This condition is satisfied by both the normal and logistic distributions.} 
The second inequality follows from similar intuition applied to individuals who do not select the investment.

\subsubsection{Odds-Based Moment Inequalities}
The conditional odds-based moment inequalities are
\begin{equation}\label{OBMI}
\begin{split}
\mathbb{E}\Bigg[
\Bigg(
S_i
\frac{1-\Phi(X_{i}\theta^*_{{\tilde{\epsilon}}}-Price_i {\gamma}^*_{{\tilde{\epsilon}}})}
{\Phi(X_{i}\theta^*_{{\tilde{\epsilon}}}-Price_i {\gamma}^*_{{\tilde{\epsilon}}})}
-(1-S_i)
\Bigg)
\Bigg|
Z_i
\Bigg] 
&\geq 0, \\
\mathbb{E}\Bigg[
\Bigg(
(1-S_i)\frac{\Phi(X_{i}\theta^*_{{\tilde{\epsilon}}}-Price_i {\gamma}^*_{{\tilde{\epsilon}}})}
{1-\Phi(X_{i}\theta^*_{{\tilde{\epsilon}}}-Price_i {\gamma}^*_{{\tilde{\epsilon}}})}
-S_i
\Bigg)
\Bigg|
Z_i
\Bigg] 
&\geq 0.
\end{split}
\end{equation}
They are derived from the unobservable conditional score equation,\hfill
\begin{equation}\label{II-Score}
\mathbb{E}\Bigg[
S_i\frac{\phi(X_{i}\beta^*_{{\tilde{\epsilon}}}-\widetilde{Price}_i {\gamma}^*_{{\tilde{\epsilon}}})}
{\Phi(X_{i}\beta^*_{{\tilde{\epsilon}}}-\widetilde{Price}_i{\gamma}^*_{{\tilde{\epsilon}}})} 
-(1-S_i)\frac{\phi(X_{i}\beta^*_{{\tilde{\epsilon}}}-\widetilde{Price}_i {\gamma}^*_{{\tilde{\epsilon}}})}
{1-\Phi(X_{i}\beta^*_{{\tilde{\epsilon}}}-\widetilde{Price}_i{\gamma}^*_{{\tilde{\epsilon}}})}
\Bigg|
X_i, \widetilde{Price}_i
\Bigg] = 0.
\end{equation}
Considering the second inequality, the score function can be rearranged to be a function of the model-predicted odds of selecting the investment,\hfill
\begin{equation}
\mathbb{E}\Bigg[
\Bigg(
(1-S_i)\frac{\Phi(X_{i}\beta^*_{{\tilde{\epsilon}}}-\widetilde{Price}_i {\gamma}^*_{{\tilde{\epsilon}}})}
{1-\Phi(X_{i}\beta^*_{{\tilde{\epsilon}}}-\widetilde{Price}_i {\gamma}^*_{{\tilde{\epsilon}}})}
-S_i
\Bigg)
\Bigg|
X_i, \widetilde{Price}_i
\Bigg] 
= 0.
\end{equation}
The advantage of this transformation is that the odds-ratio is globally convex in its arguments. Replacing the unobserved $\widetilde{Price}_i$ with $Price_i$ changes the equation into an inequality by application of Jensen's inequality due to the global convexity of the odds ratio. As the index of the odds ratio increases, the model-predicted odds of a given outcome approach positive infinity, while the odds approach zero as the index decreases. When the index is replaced with a mean-preserving spread of itself (via replacing perceived prices with prices), this first effect will usually dominate the second regardless of the distributional assumption.\footnote{Global convexity of the odds ratio is necessary for this condition to hold for all values of the index and for all magnitudes of mean-preserving spreads. This condition is satisfied by log-concave distributions, such as the normal and logistic.} This inequality holds when taking its expectation conditional on $Z_i$ by law of iterated expectations. The first inequality follows from similar intuition for those who do not select the investment.


\subsubsection{Estimation Using Moment Inequalities}
Under the information assumptions provided, the true parameters $\psi=(\theta,\sigma_{\tilde{\epsilon}})$ will be contained within the set of parameters that satisfy the inequalities, which I define as $\Psi_0$. First, because it is computationally expensive to compute the inequalities conditional on $Z_i$, I will instead use unconditional inequalities that are consistent with the conditional inequalities described above. Additionally, in small samples it is possible that the true parameters will not strictly satisfy these inequalities, so it is necessary to construct a test of the hypothesis that a given value $\psi_p=(\theta_p,\sigma_{\tilde{\epsilon},p})$ is consistent with the inequalities. To do this I employ the modified method of moments procedure described by \cite{as10}, which yields a confidence set of parameters $\hat{\Psi}_0$ that I fail to reject are consistent with the inequalities, where an element of this set is given by $(\hat{\theta}_p,\hat{\sigma}_{\tilde{\epsilon},p})$. A description of the estimation procedure is provided in Appendix \ref{appendix:MMM}.

To infer estimates of perceived returns from the estimated set of parameters that satisfy the moment inequalities, first note that, given the true $(\beta,\sigma_{\tilde{\epsilon}})$, perceived returns are given by \hfill
\begin{equation}\label{perceivedreturnsMI_true}
{Y}_i|X_i, \widetilde{Price}_i \sim \mathcal{N}(X_{i}{\beta}-\widetilde{Price}_i{},{\sigma_{\tilde{\epsilon}}}^2).
\end{equation}
Thus, even given the true $(\beta,\sigma_{\tilde{\epsilon}})$, the problem remains that we do not observe $\widetilde{Price}_i$ in the data. However, it is possible to bound perceived returns at the true parameter values using $Price_i$ and $\mathbb{E}[Price_i|Z_i]$, which we do have access to. 

For valid $Z_i$, equation (\ref{belief_restriction}) implies that it is possible to approximate $\widetilde{Price}_i$ with $\varphi Price_i+(1-\varphi)\mathbb{E}[Price_i|Z_i]$, where $\varphi$ minimizes $\mathbb{E}[\widetilde{Price}_i - (\varphi Price_i+(1-\varphi)\mathbb{E}[Price_i|Z_i])]^2$.
It must be that $\varphi \in [0,1]$, though we cannot estimate it.
Given $\varphi \in [0,1]$, bounds on perceived returns for a given $(\hat{\theta}_p,\hat{\sigma_{\tilde{\epsilon}}}_p)$ can be constructed using
\begin{equation}\label{MIdist}
{Y}_i|X_i, Price_i,Z_i \sim \mathcal{N}\big(X_{i}\hat{\theta}_p-\varphi Price_i{}-(1-\varphi)\mathbb{E}[Price_i|Z_i]{},\hat{\sigma_{\tilde{\epsilon}}}_p^2\big).
\end{equation}
Note that the PDF of this distribution is non-monotonic in $\varphi$, so setting $\varphi=0$ and $\varphi=1$ will not bound its PDF across its entire support.
Computing the distribution for all $\varphi \in [0,1]$ for each $(\hat\theta_p,\hat\sigma_{\tilde{\epsilon},p}) \in \hat\Psi_0$ is necessary to provide bounds for the perceived returns distribution.\footnote{In practice, choosing any set of values between zero and one, including zero and one, will approximate these bounds. DM describe an alternative method that can be used to bound the CDF of perceived returns.}

\section{Moment Inequality Estimation}\label{appendix:MMM}

I closely follow appendices A.5 and A.7 in \cite{dm18} to estimate the moment inequalities' confidence set for the true parameter $\psi$. Adapting DM's procedure to the current setting would account for imputation of prices. I use a simplified version of their procedure, because I assume that prices are observed for all individuals, regardless of whether they select the investment. This assumption is irrelevant to the contributions of this paper, as each method admits imputation. I also deviate from DM in how I conduct the grid search over potential parameters in order to speed computation in the absence of parallelization.

The confidence set is obtained by applying the \cite{as10} modified method of moments (MMM). This method follows the intuition of the generalized method of moments, but only penalizes moment deviations that violate the inequality while adjusting the hypothesis testing procedure to accommodate this change. I index the moment inequalities used in estimation by $\ell = 1,...,L$ and denote them
\begin{equation*}
    \bar{m}_\ell(\psi) \equiv \frac{1}{N}\sum_{i=1}^N m_{\ell}(Z_i,\psi), \qquad \ell = 1,...,L,
\end{equation*}
where $N$ is the sample size. The MMM test statistic
\begin{equation}
Q(\psi) = \sum_{\ell}^L [\min(\sqrt{N}\frac{\bar{m}_\ell(\psi)}{\hat{\sigma}_\ell(\psi)},0)]^2,
\end{equation}
gives the sum of squared inequality violations, where
\begin{equation*}
    \hat{\sigma}_\ell(\psi) = 
    \sqrt{
    \frac{1}{N}
    \sum_{i=1} ^N
    ({m}_\ell(Z_i,\psi)
    -
    \bar{m}_\ell(\psi))^2
    }.
\end{equation*}
Note that as in Section \ref{MI_method}, $X_i \subset Z_i$. $m_\ell(\cdot)$ is a conditional revealed preference or odds-based moment inequality constructed as described in DM, Appendix A.5. I compute a confidence set for the true parameter $\psi$ using the following steps, closely following DM.\\
\textbf{Step 1: define a grid $\Psi_{g}$ that overlaps with the confidence set.} I define this grid as a $K$-dimensional orthotope where $K$ is the number of scalars indexed by $k=1,...,K$ within the parameter vector $\psi$. To define this grid, I choose $\psi_{min}$ to minimize $Q(\psi)$, initializing the minimization with the control function estimates $\hat\psi_{CF} \equiv (\hat\theta_{CF},\hat\sigma_{\zeta,CF})$, which in simulations is typically near a minimum (zero) of $Q(\psi)$. The moment inequality confidence set encompass the control function estimates in simulations included later in Appendix \ref{appendix:More_Sims} when they provide consistent bounds, and there is good reason to believe that this will be the case generally (see Appendix \ref{appendix:MI_Proofs}). Because $Q(\hat\psi_{min})$ will be close to zero, it is likely to be within the 95\% confidence set, $\hat\Psi_0^{95}$, if this set is nonempty. I create boundaries in dimension $k$ by multiplying the standard error of the $k$th parameter by a large number, and adding and subtracting this value from the parameter to form bounds in the $k$th dimension.\footnote{As there are negligible computational disadvantages from having a very large initial grid, I multiply the standard errors by 20.} I repeat this for each of the $K$ parameters to obtain bounds on a $K$-dimensional initial grid $\Psi_{g}$. I fill this grid with $10 ^ K$ equidistant points.\\
\textbf{Step 2: choose a point $\psi_p \in \Psi_g$}. For speed, I test points in ascending order of their euclidean distance from $\hat\psi_{min}$. With $\psi_p$, I test the hypothesis that $\psi_p=\psi$:
\begin{equation*}
    H_0: \psi=\psi_p \qquad vs. \qquad H_0: \psi \neq \psi_p.
\end{equation*}
\textbf{Step 3: evaluate the MMM test statistic at $\psi_p$:}
\begin{equation}
Q(\psi_p) = \sum_{\ell}^L 
\big[
\min(\sqrt{N}\frac{\bar{m}_\ell(\psi_p)}{\hat{\sigma}_\ell(\psi_p)},0)
\big]
^2,
\end{equation}
\textbf{Step 4: compute the correlation matrix of the moments evaluated at $\psi_p$:}
\begin{equation*}
    \hat\Omega(\psi_p) = 
    Diag^{
    -\frac{1}{2}
    }
    (
    \hat\Sigma(\psi_p)
    )
    \hat\Sigma(\psi_p)
    Diag^{
    -\frac{1}{2}
    }
    (
    \hat\Sigma(\psi_p)
    ),
\end{equation*}
where $Diag^{    -\frac{1}{2}    }    (    \hat\Sigma(\psi_p)    )$ is the $L \times L$ diagonal matrix that shares diagonal elements with $\hat\Sigma(\psi_p)$. $Diag^{    -\frac{1}{2}    }(    \hat\Sigma(\psi_p)    )$ satisfies $Diag^{    -\frac{1}{2}    }(    \hat\Sigma(\psi_p)    )Diag^{    -\frac{1}{2}    }(    \hat\Sigma(\psi_p)    ) = Diag^{    -1    }(    \hat\Sigma(\psi_p)    )$ where 
\begin{equation*}
    \hat\Sigma(\psi_p) = 
    \frac{1}{N}
    \sum_{i=1}^N
    (
    m(Z_i,\psi_p)
    -
    \bar m(\psi_p)
    )
    (
    m(Z_i,\psi_p)
    -
    \bar m(\psi_p)
    )',
\end{equation*}
$m(Z_i,\psi_p) = ( m_1(Z_i,\psi_p),...,m_L(Z_i,\psi_p))$,
and $\bar m(\psi_p) = (\bar m_1(\psi_p),...,\bar m_L(\psi_p))$, where
\begin{equation*}
    \bar{m}_\ell(\psi_p) \equiv \frac{1}{N}\sum_{i=1}^N m_{\ell}(Z_i,\psi_p), \qquad \forall \ell = 1,...,L.
\end{equation*}
\textbf{Step 5: simulate the asymptotic distribution of $Q(\psi_p)$.} Take $R=1000$ draws from the multivariate normal distribution $\mathcal{N}(0_L,I_L)$ where $0_L$ is a vector of zeros and $I_L$ is an $L$-dimensional identity matrix. Denote each of these draws as $\chi_r$. Define the criterion function $Q_{N,r}^{AA}(\psi_p)$ as 
\begin{equation*}
    Q_{N,r}^{AA}(\psi_p) = 
    \sum_{\ell=1}^L
    \Bigg[
    \bigg(
    \min
    \Big(
    [
    \hat\Omega
    ^{
    \frac{1}{2}
    }
    (\psi_p)
    \chi_r
    ]
    _\ell
    ,0
    \Big)
    \bigg)    ^2
    \times
    \mathbbm{1}
    \bigg(
    \sqrt{N}
    \frac
    {    \bar{m}_\ell(\psi_p)}
    {\hat{\sigma}_\ell(\psi_p)}
    \leq 
    \sqrt{
    \ln N}
    \bigg)
    \bigg]   ,
    \end{equation*}
where $[
    \hat\Omega
    ^{
    \frac{1}{2}
    }
    (\psi_p)
    \chi_r
    ]
    _\ell$
is the $\ell th$ element of the vector $\hat\Omega
    ^{
    \frac{1}{2}
    }
    (\psi_p)
    \chi_r$.\\
\textbf{Step 6: compute the critical value.} The critical value $\hat c _N ^{AA}(\psi_p,1-\alpha)$ is the $(1-\alpha)$-quantile distribution of the distribution of $Q_{N,r}^{AA}(\psi_p)$ across the $R$ draws taken in step 5.\\
\textbf{Step 7: reject or fail to reject $\psi_p$.} If $Q(\psi_p) \leq \hat c _N ^{AA}(\psi_p,1-\alpha)$, include $\psi_p$ in the estimated $(1-\alpha)\%$ confidence set, $\hat\Psi^{1-\alpha}$ and the (initially empty) grid $\Psi_{g'}$ that will contain the confidence set.\\
\textbf{Step 8: repeat steps 2 through 7 until a $\psi_p$ is not rejected.} This will likely occur at the first point checked, $\psi_{min}$, as this parameter minimizes $Q(\psi_p)$, though it does not maximize $\hat c _N ^{AA}(\psi_p,1-\alpha)$.\\
\textbf{Step 9: form a small grid around each $\psi_p$ in $\hat\Psi^{1-\alpha}$.} Form $\Psi_{g,p}$, a local $K$-dimensional orthotope with 3 equidistant points in each dimension (with distance between points defined as in step 1), centered around $\psi_p$ for each $\psi_p$ in $\hat\Psi^{1-\alpha}$. Add $\Psi_{g,p}$ to the grid $\Psi_{g'}$ that will contain the confidence set. \\
\textbf{Step 10: repeat steps 3 through 7 for every point in $\Psi_{g'}$ that has not yet been checked.}\\
\textbf{Step 11: iterate on steps 9 and 10 until all points in $\Psi_{g'}$ have been checked.}\\
\textbf{Step 12: ensure desired grid fineness.} If the number of elements of the set $\hat\Psi^{1-\alpha}$ is below the desired minimum number, set the distance between grid points at one-half of the current value and repeat step 11. Repeat this step until the number of elements of $\hat\Psi^{1-\alpha}$ exceeds the desired number of such elements.

\section{Moment Inequalities and Endogeneity}\label{appendix:MI_Proofs}

\normalsize

For proofs of the validity of the moment inequalities for providing a confidence set that consistently bounds the true parameter vector, $(\theta,\sigma_\epsilon)$, in the context of the setting presented in Section \ref{MI_method}, see DM. The inequalities also appear to consistently bound perceived returns in simulations when there is correlation between perceived prices and the unobserved error in perceived returns and correlation between information frictions and the unobserved error in perceived returns under the assumption $\frac{\rho}{1-\rho}\geq0$, which is weaker than the assumption described in Section \ref{MI_method}. I provide proofs of consistency here for the revealed preference moment inequalities, and arguments for consistency for the odds-based moment inequalities, borrowing from the proofs provided by DM. Note that the parameters relevant to this section are $(\theta,\sigma_\xi)$, not those used in Section \ref{MI_method}. I use the notation $(\theta^*_\xi,\gamma^*_\xi, \rho^*_\xi)=\big(\frac{\theta}{\sigma_\xi}, \frac{1}{\sigma_\xi},\frac{\rho}{\sigma_\xi}\big)$ throughout the following while assuming
\begin{equation}\label{error_MI_endo}
    \xi_i|X_i,Price_i,u_i \sim \mathcal{N}(0,\sigma_\xi^2).
\end{equation}

The condition $\frac{\rho}{1-\rho}\geq0$ entails a setting in which both endogeneity and measurement error work against the causal effect of price on selection, yet the causal effect dominates, producing attenuation bias in estimates of the effect of perceived prices on selection if misperceptions and endogeneity are ignored. 
Heuristically, this restriction suggests that including $u_i$ with a multiplier of $-1$ via prices (recalling the definition of prices given in (\ref{Price_Z})) will strengthen the overall effect of prices on inequalities derived from (\ref{Perceived_Returns_CF}) relative to the alternative of multiplying this error (or an estimate of it) by $(-1+\rho)$.

Note that $\frac{\rho}{1-\rho}\geq0$ is equivalent to $\rho \in [0,1]$. The assumption that $\rho \geq 0$ seems plausible, prices may be higher for individuals for whom perceived returns for selection are higher due to higher demand. The additional assumption that $\rho \leq 1$ has no such obvious theoretical support. Under this condition the moment inequalities appear to provide consistent bounds for $\sigma_\xi$, but not necessarily $\sigma$. As these parameters serve the same function, this has no effect on the predictive capacity of any resulting estimates of perceived returns.

I begin by presenting a lemma that will be useful in the subsequent proofs. It also serves as the main point of departure from the proofs provided by DM.
\begin{lem}
If equations $(\ref{Selection})$, $(\ref{perceivedreturns})$, $(\ref{Price_Z})$, and $(\ref{Perceived_Returns_CF})$ hold and $\frac{\rho}{1-\rho}\geq0$, then 
\begin{equation}\label{error_inequality}
\mathbb{E}[u_i(1-\rho)|S_i=0,Z_i]
\geq
0
\geq 
\mathbb{E}[u_i(1-\rho)|S_i=1,Z_i].
\end{equation}
\end{lem}
\noindent \textbf{Proof:} From the definition of $S_i$ given in (\ref{Selection}) and (\ref{perceivedreturns}), substituting in the expression for perceived returns in (\ref{Perceived_Returns_CF}) implies
\begin{equation}\label{unuS0main}
\begin{gathered}
\mathbb{E}[
u_i(1-\rho)
|
S_i=0
,
Z_i
]\\
=
\mathbb{E}[
u_i(1-\rho)
|
X_i\theta -Price_i +u_i\rho+\xi_i \leq0
,
Z_i
].
\end{gathered}
\end{equation}
Substituting in the definition of $Price_i$ provided in (\ref{Price_Z}) and rearranging the conditioning inequality implies
\begin{equation}
\begin{split}
&\mathbb{E}[
u_i(1-\rho)
|
X_i\theta -Price_i +u_i\rho+\xi_i \leq0
,
Z_i
] \\
=
&\mathbb{E}[
u_i(1-\rho)
|
X_i\theta -Z_i\delta -u_i(1-\rho)+\xi_i \leq0
,
Z_i
] \\
=
&\mathbb{E}[
u_i(1-\rho)
|
u_i(1-\rho)
\geq 
(X_i\theta-Z_i\delta+\xi_i)
,
Z_i
].
\end{split}
\end{equation}
Given the property of expectations of truncated variables that $\mathbb{E}[X|X \geq Y]\geq \mathbb{E}[X]$, it follows that 
\begin{equation}\label{unuS0}
\begin{split}
&\mathbb{E}[
u_i(1-\rho)
|
u_i(1-\rho)
\geq 
(X_i\theta-Z_i\delta+\xi_i)
,
Z_i
]\\
\geq 
&\mathbb{E}[
u_i(1-\rho)
|
Z_i
]\\
=& 0,
\end{split}
\end{equation}
where the last equality follows from the definition of $u_i$ given in (\ref{Price_Z}). The definition of $S_i$ given in (\ref{Selection}) and (\ref{perceivedreturns}), substituting in the expression of perceived returns in (\ref{Perceived_Returns_CF}), also implies
\begin{equation}\label{unuS1main}
\begin{gathered}
\mathbb{E}[
u_i(1-\rho)
|
S_i=1
,
Z_i
]\\
=
\mathbb{E}[
u_i(1-\rho)
|
X_i\theta -Price_i +u_i\rho+\xi_i \geq0
,
Z_i
].
\end{gathered}
\end{equation}
Substituting in the definition of $Price_i$ provided in (\ref{Price_Z}) and rearranging the conditioning inequality implies
\begin{equation}
\begin{split}
&\mathbb{E}[
u_i(1-\rho)
|
X_i\theta -Price_i +u_i\rho+\xi_i \geq0
,
Z_i
] \\
=
&\mathbb{E}[
u_i(1-\rho)
|
X_i\theta -Z_i\delta -u_i(1-\rho)+\xi_i \geq0
,
Z_i
] \\
=
&\mathbb{E}[
u_i(1-\rho)
|
u_i(1-\rho)
\leq 
(X_i\theta-Z_i\delta+\xi_i)
,
Z_i
].
\end{split}
\end{equation}
Given the property of expectations of truncated variables that $\mathbb{E}[X|X \leq Y]\leq \mathbb{E}[X]$, it follows that 
\begin{equation}\label{unuS1}
\begin{split}
&\mathbb{E}[
u_i(1-\rho)
|
u_i(1-\rho)
\leq 
(X_i\theta-Z_i\delta+\xi_i)
,
Z_i
]\\
\leq 
&\mathbb{E}[
u_i(1-\rho)
|
Z_i
]\\
=& 0,
\end{split}
\end{equation}
where the last equality follows from the definition of $u_i$ given in (\ref{Price_Z}). Substituting (\ref{unuS0}) into (\ref{unuS0main}) and (\ref{unuS1}) into (\ref{unuS1main}) implies (\ref{error_inequality}). $\blacksquare$



\subsection{Proof of Revealed Preference Inequality Robustness to Endogeneity}
\begin{lem}
Suppose equations $(\ref{Selection})$, $(\ref{perceivedreturns})$, and $(\ref{Perceived_Returns_CF})$ hold. Then 
\begin{equation}\label{selection_rp1}
\mathbb{E}\Bigg[
S_i 
\big(
X_i\theta-{Price_i}+u_i\rho+\xi_i
\big)
\Bigg|Z_i\Bigg]
\geq 0.
\end{equation}
\end{lem}
\noindent \textbf{Proof:} From equations $(\ref{Selection})$, $(\ref{perceivedreturns})$, and $(\ref{Perceived_Returns_CF})$,
\begin{equation}
S_i = \mathbbm{1}\{X_i\theta-{Price_i}+u_i\rho+\xi_i
\geq 0\}.
\end{equation}
This implies 
\begin{equation}
    S_i
    \big(
    X_i\theta-{Price_i}+u_i\rho+\xi_i
    \big) 
    \geq 0.
\end{equation}
This inequality holds for every individual $i$, therefore it will hold in expectation conditional on 
$Z_i$. $\blacksquare$

\begin{lem}
Equations $(\ref{Selection})$, $(\ref{perceivedreturns})$, $(\ref{Price_Z})$, $(\ref{Perceived_Returns_CF})$, and $(\ref{error_MI_endo})$ imply that
\begin{equation}\label{IM_rp1}
\begin{gathered}
\mathbb{E}\Bigg[
S_i 
\big(
X_i\theta^*_\xi-{Z_i}\delta\gamma^*_\xi
\big)
\\
+
(1-S_i)
\Bigg(
u_i(\gamma^*_\xi-\rho^*_\xi)
+
\frac
{\phi\big(X_i\theta^*_\xi-Price_i\gamma^*_\xi+u_i\rho^*_\xi\big)}
{1-\Phi\big(X_i\theta^*_\xi-Price_i\gamma^*_\xi+u_i\rho^*_\xi\big)}
\Bigg)
\Bigg|
Z_i
\Bigg]
\geq 0
\end{gathered}
\end{equation}
\end{lem}
\noindent \textbf{Proof:} Equation (\ref{selection_rp1}) and the definition of $Price_i$ from equation (\ref{Price_Z}) imply
\begin{equation}\label{rp1_splitE}
\mathbb{E}
[
S_i 
(
X_i\theta-{Z_i}\delta
)
|
Z_i
]
-
\mathbb{E}
[
S_iu_i(1-\rho)
|
Z_i
\big]
+
\mathbb{E}
[
S_i 
\xi_i
|
Z_i
]
\geq 0.
\end{equation}
The assumption in (\ref{Price_Z}) implies that $\mathbb{E}[u_i|Z_i]=0$, so it follows that
\begin{equation*}
\mathbb{E}[S_i
u_i(1-\rho)
+
(1-S_i)
u_i(1-\rho)
|Z_i]=0.
\end{equation*}
Expression (\ref{error_MI_endo}) implies that $\mathbb{E}[\xi_i|X_i,Price_i,u_i]=0$, which implies
\begin{equation*}
\mathbb{E}[S_i\xi_i
+
(1-S_i)\xi_i
|X_i,Price_i,u_i]
=0.
\end{equation*}
Assuming the distribution of $Z_i$ conditional on $X_i,Price_i,u_i$ is degenerate and applying the law of iterated expectations, the preceding two equations allow us to rewrite equation (\ref{rp1_splitE}) as
\begin{equation}\label{rp1_2}
\mathbb{E}
[
S_i 
(
X_i\theta-{Z_i}\delta
)
|
Z_i
]
+
\mathbb{E}
[
(1-S_i)
u_i
(1-\rho)
|
Z_i
]
-
\mathbb{E}\big[
(1-S_i)
\xi_i
\big|
Z_i
\big]
\geq 0.
\end{equation}
Assuming the distribution of $Z_i$ conditional on $X_i,Price_i,u_i$ is degenerate and applying the law of iterated expectations also implies
\begin{equation*}
\begin{split}
\mathbb{E}[(1-S_i)\xi_i|Z_i] &= 
\mathbb{E}\big[\mathbb{E}[(1-S_i)\xi_i|S_i,X_i,Price_i,u_i]\big|Z_i\big] 
\\
&=
\mathbb{E}\big[
\mathbb{E}[
(1-S_i)|X_i,Price_i,u_i]\mathbb{E}[\xi_i|S_i,X_i,Price_i,u_i]
\big|Z_i\big] 
\\
&=
\mathbb{E}\big[
P\big(S_i=1|X_i,Price_i,u_i\big)\times 0 \times \mathbb{E}[\xi_i|S_i=1,X_i,Price_i,u_i] 
\\
&\indent + 
P\big(S_i=0|X_i,Price_i,u_i\big)\times 1 \times \mathbb{E}[\xi_i|S_i=0,X_i,Price_i,u_i]
\big|Z_i\big] 
\\
&=
\mathbb{E}\big[
P\big(S_i=0|X_i,Price_i,u_i\big)\mathbb{E}[\xi_i|S_i=0,X_i,Price_i,u_i] 
\big|Z_i\big] 
\\
&=
\mathbb{E}\big[
\mathbb{E}[(1-S_i)|X_i,Price_i,u_i]\mathbb{E}[\xi_i|S_i=0,X_i,Price_i,u_i]
\big|Z_i\big] 
\\
&=
\mathbb{E}\Big[
\mathbb{E}\big[(1-S_i)
\mathbb{E}[\xi_i|S_i=0,X_i,Price_i,u_i]
\big|X_i,Price_i,u_i]
\Big|Z_i\Big] 
\\
&=
\mathbb{E}\big[(1-S_i)
\mathbb{E}[\xi_i|S_i=0,X_i,Price_i,u_i]
\big|Z_i].
\end{split}
\end{equation*}
This allows us to rewrite equation (\ref{rp1_2}) as 
\begin{equation}\label{rp1_rearrange}
\mathbb{E}
\big
[
S_i 
(
X_i\theta-{Z_i}\delta
)
+
(1-S_i)
\big(
u_i
(1-\rho)
-
\mathbb{E}
[
\xi_i|S_i=0,X_i,Price_i,u_i
]
\big)
\big|
Z_i
\big]
\geq 0.
\end{equation}
Using the definition of $S_i$ from equation (\ref{Selection}) and substituting in equations (\ref{perceivedreturns}) and (\ref{Perceived_Returns_CF}), it follows that
\begin{equation*}
\begin{split}
\mathbb{E}[\xi_i|S_i=0,X_i,Price_i,u_i]
=
\mathbb{E}\big[
&
\xi_i
\big|
\big(
-\xi_i \geq X_i\theta-{Price_i}+u_i\rho
\big)
,X_i,Price_i,u_i\big]
\\
=
-\mathbb{E}\big[-
&
\xi_i
\big|
\big(
-\xi_i \geq X_i\theta-{Price_i}+u_i\rho
\big)
,X_i,Price_i,u_i\big],
\end{split}
\end{equation*}
which allows us to rewrite
\begin{equation}
\mathbb{E}[\xi_i|S_i=0,X_i,Price_i,u_i]
=
-\sigma_\xi
\frac{
\phi\big(X_i\theta^*_\xi-{Price_i}\gamma^*_\xi+u_i\rho^*_\xi\big)
}
{
1-\Phi\big(X_i\theta^*_\xi-{Price_i}\gamma^*_\xi+u_i\rho^*_\xi\big)
}
\end{equation}
using Expression (\ref{error_MI_endo}) and applying the symmetry of the normal distribution. Equation (\ref{IM_rp1}) follows by applying this equality to (\ref{rp1_rearrange}) and dividing each side of the resulting inequality by $\sigma_\xi$. $\blacksquare$

\begin{lem}
Given $\frac{\rho}{1-\rho}\geq 0$, equations $(\ref{Selection})$, $(\ref{perceivedreturns})$, $(\ref{Price_Z})$, and $(\ref{Perceived_Returns_CF})$ imply
\begin{equation}\label{rp1_drop_uv}
\begin{gathered}
\mathbb{E}\Bigg[
(1-S_i)
\Bigg(
u_i\gamma^*_\xi
+
\frac
{\phi\big(X_i\theta^*_\xi-Price_i\gamma^*_\xi\big)}
{1-\Phi\big(X_i\theta^*_\xi-Price_i\gamma^*_\xi\big)}
\Bigg)
\Bigg|
Z_i
\Bigg]\\
\geq\\
\mathbb{E}\Bigg[
(1-S_i)
\Bigg(
u_i(\gamma^*_\xi-\rho^*_\xi)
+
\frac
{\phi\big(X_i\theta^*_\xi-Price_i\gamma^*_\xi+u_i\rho^*_\xi\big)}
{1-\Phi\big(X_i\theta^*_\xi-Price_i\gamma^*_\xi+u_i\rho^*_\xi\big)}
\Bigg)
\Bigg|
Z_i
\Bigg]
\end{gathered}
\end{equation}
\end{lem}
\noindent \textbf{Proof:} Using the definition of $Price_i$ from equation $(\ref{Price_Z})$, it follows that
\begin{equation}
\begin{gathered}
\mathbb{E}\Bigg[
(1-S_i)
\Bigg(
u_i(\gamma^*_\xi-\rho^*_\xi)
+
\frac
{\phi\big(X_i\theta^*_\xi-Price_i\gamma^*_\xi+u_i\rho^*_\xi\big)}
{1-\Phi\big(X_i\theta^*_\xi-Price_i\gamma^*_\xi+u_i\rho^*_\xi\big)}
\Bigg)
\Bigg|
Z_i
\Bigg]\\
=\\
\mathbb{E}\Bigg[
(1-S_i)
\Bigg(
u_i(\gamma^*_\xi-\rho^*_\xi)
+
\frac
{\phi\big(X_i\theta^*_\xi-Z_i\delta\gamma^*_\xi-u_i(\gamma^*_\xi-\rho^*_\xi)\big)}
{1-\Phi\big(X_i\theta^*_\xi-Z_i\delta\gamma^*_\xi-u_i(\gamma^*_\xi-\rho^*_\xi)\big)}
\Bigg)
\Bigg|
Z_i
\Bigg].
\end{gathered}
\end{equation}
The law of iterated expectations and $S_i\in \{0,1\}$ implies
\begin{equation}\label{rp1_error_add_S0Z}
\begin{gathered}
\mathbb{E}\Bigg[
(1-S_i)
\Bigg(
u_i(\gamma^*_\xi-\rho^*_\xi)
+
\frac
{\phi\big(X_i\theta^*_\xi-Z_i\delta\gamma^*_\xi-u_i(\gamma^*_\xi-\rho^*_\xi)\big)}
{1-\Phi\big(X_i\theta^*_\xi-Z_i\delta\gamma^*_\xi-u_i(\gamma^*_\xi-\rho^*_\xi)\big)}
\Bigg)
\Bigg|
Z_i
\Bigg]\\
=\\
\mathbb{E}\Bigg[
(1-S_i)
\mathbb{E}
\bigg[
u_i(\gamma^*_\xi-\rho^*_\xi)
+
\frac
{\phi\big(X_i\theta^*_\xi-Z_i\delta\gamma^*_\xi-u_i(\gamma^*_\xi-\rho^*_\xi)\big)}
{1-\Phi\big(X_i\theta^*_\xi-Z_i\delta\gamma^*_\xi-u_i(\gamma^*_\xi-\rho^*_\xi)\big)}
\Bigg)
\bigg|
S_i=0,Z_i
\bigg]
\Bigg|
Z_i
\Bigg].
\end{gathered}
\end{equation}
Because
\begin{equation*}
\frac{
\partial \frac
{\phi(-x)}
{1-\Phi(-x)}}
{\partial x}
\in (-1,0),
\end{equation*}
it follows that the expression
\begin{equation}
u_i(\gamma^*_\xi-\rho^*_\xi)
+
\frac
{\phi\big(X_i\theta^*_\xi-Z_i\delta\gamma^*_\xi-u_i(\gamma^*_\xi-\rho^*_\xi)\big)}
{1-\Phi\big(X_i\theta^*_\xi-Z_i\delta\gamma^*_\xi-u_i(\gamma^*_\xi-\rho^*_\xi)\big)}
\end{equation}
is monotonically increasing in $u_i(\gamma^*_\xi-\rho^*_\xi)$. It follows then that adding a positive value to this value will increase the value of the function. From (\ref{error_inequality}) and the condition $\frac{\rho^*_\xi}{\gamma^*_\xi-\rho^*_\xi} \geq 0$, $\frac{\rho^*_\xi}{\gamma^*_\xi-\rho^*_\xi}\mathbb{E}[u_i(\gamma^*_\xi-\rho^*_\xi)|S_i=0,Z_i] \geq 0$ $\forall i$, so it follows that
\begin{equation}\label{rp1_add_constantS0}
\begin{gathered}
\mathbb{E}
\bigg
[
u_i(\gamma^*_\xi-\rho^*_\xi)+\mathbb{E}[u_i\rho^*_\xi|S_i=0,Z_i]
\\+
\frac
{\phi\big(X_i\theta^*_\xi-Z_i\delta\gamma^*_\xi-u_i(\gamma^*_\xi-\rho^*_\xi)-\mathbb{E}[u_i\rho^*_\xi|S_i=0,Z_i]\big)}
{1-\Phi\big(X_i\theta^*_\xi-Z_i\delta\gamma^*_\xi-u_i(\gamma^*_\xi-\rho^*_\xi)-\mathbb{E}[u_i\rho^*_\xi|S_i=0,Z_i]\big)}
\bigg
|
S_i=0,Z_i
\bigg
]
\\
=
\\
\mathbb{E}
\bigg
[
\big(
u_i(\gamma^*_\xi-\rho^*_\xi)
+
\frac{\rho^*_\xi}{\gamma^*_\xi-\rho^*_\xi}
\mathbb{E}[u_i(\gamma^*_\xi-\rho^*_\xi)|S_i=0,Z_i]
\big)
\\
+
\frac
{\phi\Big(X_i\theta^*_\xi-Z_i\delta\gamma^*_\xi-\big(u_i(\gamma^*_\xi-\rho^*_\xi)+
\frac{\rho^*_\xi}{\gamma^*_\xi-\rho^*_\xi}
\mathbb{E}[u_i(\gamma^*_\xi-\rho^*_\xi)|S_i=0,Z_i]\big)\Big)}
{1-\Phi\Big(X_i\theta^*_\xi-Z_i\delta\gamma^*_\xi-\big(u_i(\gamma^*_\xi-\rho^*_\xi)+
\frac{\rho^*_\xi}{\gamma^*_\xi-\rho^*_\xi}
\mathbb{E}[u_i(\gamma^*_\xi-\rho^*_\xi)|S_i=0,Z_i]\big)\Big)}
\bigg
|
S_i=0,Z_i
\bigg
]
\\
\geq \\
\mathbb{E}
\bigg
[
u_i(\gamma^*_\xi-\rho^*_\xi)
+
\frac
{\phi\big(X_i\theta^*_\xi-Z_i\delta\gamma^*_\xi-u_i(\gamma^*_\xi-\rho^*_\xi)\big)}
{1-\Phi\big(X_i\theta^*_\xi-Z_i\delta\gamma^*_\xi-u_i(\gamma^*_\xi-\rho^*_\xi)\big)}
\bigg
|
S_i=0,Z_i
\bigg
],
\end{gathered}
\end{equation}
where the second line relates to the third by this addition, and the first relates to the second by algebraic simplifications. Finally, because the term
\begin{equation*}
    u_i(\gamma^*_\xi-\rho^*_\xi)+\mathbb{E}[u_i\rho^*_\xi|S_i=0,Z_i]
\end{equation*}
and the term 
\begin{equation*}
    \frac
{\phi\big(X_i\theta^*_\xi-Z_i\delta\gamma^*_\xi-u_i(\gamma^*_\xi-\rho^*_\xi)-\mathbb{E}[u_i\rho^*_\xi|S_i=0,Z_i]\big)}
{1-\Phi\big(X_i\theta^*_\xi-Z_i\delta\gamma^*_\xi-u_i(\gamma^*_\xi-\rho^*_\xi)-\mathbb{E}[u_i\rho^*_\xi|S_i=0,Z_i]\big)}
\end{equation*}
are globally convex in $-u_i$, the entire function is globally convex in $-u_i$. It follows that
\begin{equation}
\begin{gathered}
\mathbb{E}
\bigg[
u_i\gamma^*_\xi+
\frac
{\phi\big(X_i\theta^*_\xi-Z_i\delta\gamma^*_\xi-u_i\gamma^*_\xi\big)}
{1-\Phi\big(X_i\theta^*_\xi-Z_i\delta\gamma^*_\xi-u_i\gamma^*_\xi\big)}
\bigg
|S_i=0,Z_i
\bigg
]
\\
\geq \\
\mathbb{E}
\bigg[
u_i(\gamma^*_\xi-\rho^*_\xi)+\mathbb{E}[u_i\rho^*_\xi|S_i=0,Z_i]
\\+
\frac
{\phi\big(X_i\theta^*_\xi-Z_i\delta\gamma^*_\xi-u_i(\gamma^*_\xi-\rho^*_\xi)-\mathbb{E}[u_i\rho^*_\xi|S_i=0,Z_i]\big)}
{1-\Phi\big(X_i\theta^*_\xi-Z_i\delta\gamma^*_\xi-u_i(\gamma^*_\xi-\rho^*_\xi)-\mathbb{E}[u_i\rho^*_\xi|S_i=0,Z_i]\big)}
\bigg
|S_i=0,Z_i
\bigg
]
\end{gathered}
\end{equation}
by Jensen's inequality. Combining this inequality with that in (\ref{rp1_add_constantS0}) yields the result 
\begin{equation}
\begin{gathered}
\mathbb{E}
\bigg[
u_i\gamma^*_\xi+
\frac
{\phi\big(X_i\theta^*_\xi-Z_i\delta\gamma^*_\xi-u_i\gamma^*_\xi\big)}
{1-\Phi\big(X_i\theta^*_\xi-Z_i\delta\gamma^*_\xi-u_i\gamma^*_\xi\big)}
\bigg
|S_i=0,Z_i
\bigg
]
\\
\geq \\
\mathbb{E}
\bigg
[
u_i(\gamma^*_\xi-\rho^*_\xi)
+
\frac
{\phi\big(X_i\theta^*_\xi-Z_i\delta\gamma^*_\xi-u_i(\gamma^*_\xi-\rho^*_\xi)\big)}
{1-\Phi\big(X_i\theta^*_\xi-Z_i\delta\gamma^*_\xi-u_i(\gamma^*_\xi-\rho^*_\xi)\big)}
\bigg
|
S_i=0,Z_i
\bigg
].
\end{gathered}
\end{equation}
It follows immediately that 
\begin{equation}
\begin{gathered}
\mathbb{E}
\Bigg[
(1-S_i)
\mathbb{E}
\bigg[
u_i\gamma^*_\xi+
\frac
{\phi\big(X_i\theta^*_\xi-Z_i\delta\gamma^*_\xi-u_i\gamma^*_\xi\big)}
{1-\Phi\big(X_i\theta^*_\xi-Z_i\delta\gamma^*_\xi-u_i\gamma^*_\xi\big)}
\bigg
|S_i=0,Z_i
\bigg
]
\Bigg
|Z_i
\Bigg
]
\\
\geq \\
\mathbb{E}
\Bigg
[
(1-S_i)
\mathbb{E}
\bigg[
u_i(\gamma^*_\xi-\rho^*_\xi)
+
\frac
{\phi\big(X_i\theta^*_\xi-Z_i\delta\gamma^*_\xi-u_i(\gamma^*_\xi-\rho^*_\xi)\big)}
{1-\Phi\big(X_i\theta^*_\xi-Z_i\delta\gamma^*_\xi-u_i(\gamma^*_\xi-\rho^*_\xi)\big)}
\Bigg
|
S_i=0,Z_i
\bigg
]
\Bigg
|Z_i
\Bigg
].
\end{gathered}
\end{equation}
Equation (\ref{rp1_drop_uv}) follows from this by substituting in the definition of $Price_i$ from (\ref{Price_Z}) and applying the law of iterated expectations. $\blacksquare$

\noindent\textbf{Corollary 1} \textit{Given (\ref{IM_rp1}), (\ref{rp1_drop_uv}), and the definition of $Price_i$ given in $(\ref{Price_Z})$, it follows that}
\begin{equation}\label{cor1_1}
\mathbb{E}\Bigg[
S_i (X_i\theta^*_\xi-Price_i\gamma^*_\xi)+
(1-S_i)
\frac
{\phi(X_i\theta^*_\xi-Price_i\gamma^*_\xi)}
{1-\Phi(X_i\theta^*_\xi-Price_i\gamma^*_\xi)}
\Bigg|Z_i\Bigg]
\geq 0.
\end{equation}

\noindent\textbf{Proof:} The result follows from equations (\ref{IM_rp1}), (\ref{rp1_drop_uv}), substituting $\mathbb{E}[-S_iu_i=(1-S_i)u_i|Z_i]$, and substituting in the definition of $Price_i$ given in $(\ref{Price_Z})$. $\blacksquare$

\begin{lem}
Suppose equations $(\ref{Selection})$, $(\ref{perceivedreturns})$, and $(\ref{Perceived_Returns_CF})$ hold. Then 
\begin{equation}\label{selection_rp2}
\mathbb{E}\Bigg[
-(1-S_i)
\big(
X_i\theta-{Price_i}+u_i\rho+\xi_i
\big)
\Bigg|Z_i\Bigg]
\geq 0.
\end{equation}
\end{lem}
\noindent \textbf{Proof:} From equations $(\ref{Selection})$, $(\ref{perceivedreturns})$, and $(\ref{Perceived_Returns_CF})$,
\begin{equation}
S_i = \mathbbm{1}\{X_i\theta-{Price_i}+u_i\rho+\xi_i
\geq 0\}.
\end{equation}
This implies 
\begin{equation}
    -(1-S_i)
    \big(
    X_i\theta-{Price_i}+u_i\rho+\xi_i
    \big) 
    \geq 0.
\end{equation}
This inequality holds for every individual $i$, therefore it will hold in expectation conditional on 
$Z_i$. $\blacksquare$

\begin{lem}
Equations $(\ref{Selection})$, $(\ref{perceivedreturns})$, $(\ref{Price_Z})$, $(\ref{Perceived_Returns_CF})$, and $(\ref{error_MI_endo})$ imply that
\begin{equation}\label{IM_rp2}
\begin{gathered}
\mathbb{E}\Bigg[
-(1-S_i)
\big(
X_i\theta^*_\xi-{Z_i}\delta\gamma^*_\xi
\big)
\\
+
S_i
\Bigg(-
u_i(\gamma^*_\xi-\rho^*_\xi)
+
\frac
{\phi\big(X_i\theta^*_\xi-Price_i\gamma^*_\xi+u_i\rho^*_\xi\big)}
{\Phi\big(X_i\theta^*_\xi-Price_i\gamma^*_\xi+u_i\rho^*_\xi\big)}
\Bigg)
\Bigg|
Z_i
\Bigg]
\geq 0
\end{gathered}
\end{equation}
\end{lem}
\noindent \textbf{Proof:} Equation (\ref{selection_rp2}) and the definition of $Price_i$ from equation (\ref{Price_Z}) imply
\begin{equation}\label{rp2_splitE}
\mathbb{E}
[-
(1-S_i)
(
X_i\theta-{Z_i}\delta
)
|
Z_i
]
+
\mathbb{E}
[
(1-S_i)
u_i(1-\rho)
|
Z_i
\big]
-
\mathbb{E}
[
(1-S_i)
\xi_i
|
Z_i
]
\geq 0.
\end{equation}
The assumption in (\ref{Price_Z}) implies that $\mathbb{E}[u_i|Z_i]=0$, so it follows that
\begin{equation*}
\mathbb{E}[S_i
u_i(1-\rho)
+
(1-S_i)
u_i(1-\rho)
|Z_i]=0.
\end{equation*}
Expression (\ref{error_MI_endo}) implies that $\mathbb{E}[\xi_i|X_i,Price_i,u_i]=0$, which implies
\begin{equation*}
\mathbb{E}[S_i\xi_i
+
(1-S_i)\xi_i
|X_i,Price_i,u_i]
=0.
\end{equation*}
Assuming the distribution of $Z_i$ conditional on $X_i,Price_i,u_i$ is degenerate and applying the law of iterated expectations, the preceding two equations allow us to rewrite equation (\ref{rp2_splitE}) as
\begin{equation}\label{rp2_2}
\mathbb{E}
[-
(1-S_i)
(
X_i\theta-{Z_i}\delta
)
|
Z_i
]
-
\mathbb{E}
[
S_i
u_i
(1-\rho)
|
Z_i
]
+
\mathbb{E}\big[
S_i
\xi_i
\big|
Z_i
\big]
\geq 0.
\end{equation}
Assuming the distribution of $Z_i$ conditional on $X_i,Price_i,u_i$ is degenerate and applying the law of iterated expectations also implies
\begin{equation*}
\begin{split}
\mathbb{E}[S_i\xi_i|Z_i] &= 
\mathbb{E}\big[\mathbb{E}[S_i\xi_i|S_i,X_i,Price_i,u_i]\big|Z_i\big] 
\\
&=
\mathbb{E}\big[
\mathbb{E}[
S_i|X_i,Price_i,u_i]\mathbb{E}[\xi_i|S_i,X_i,Price_i,u_i]
\big|Z_i\big] 
\\
&=
\mathbb{E}\big[
P\big(S_i=1|X_i,Price_i,u_i\big)\times 1 \times \mathbb{E}[\xi_i|S_i=1,X_i,Price_i,u_i] 
\\
&\indent + 
P\big(S_i=0|X_i,Price_i,u_i\big)\times 0 \times \mathbb{E}[\xi_i|S_i=0,X_i,Price_i,u_i]
\big|Z_i\big] 
\\
&=
\mathbb{E}\big[
P\big(S_i=1|X_i,Price_i,u_i\big)\mathbb{E}[\xi_i|S_i=1,X_i,Price_i,u_i] 
\big|Z_i\big] 
\\
&=
\mathbb{E}\big[
\mathbb{E}[S_i|X_i,Price_i,u_i]\mathbb{E}[\xi_i|S_i=1,X_i,Price_i,u_i]
\big|Z_i\big] 
\\
&=
\mathbb{E}\Big[
\mathbb{E}\big[
S_i
\mathbb{E}[\xi_i|S_i=1,X_i,Price_i,u_i]
\big|X_i,Price_i,u_i]
\Big|Z_i\Big] 
\\
&=
\mathbb{E}\big[
S_i
\mathbb{E}[\xi_i|S_i=1,X_i,Price_i,u_i]
\big|Z_i].
\end{split}
\end{equation*}
This allows us to rewrite equation (\ref{rp2_2}) as 
\begin{equation}\label{rp2_rearrange}
\mathbb{E}
\big
[-
(1-S_i)
(
X_i\theta-{Z_i}\delta
)
+
S_i
\big(
-
u_i
(1-\rho)
+
\mathbb{E}
[
\xi_i|S_i=1,X_i,Price_i,u_i
]
\big)
\big|
Z_i
\big]
\geq 0.
\end{equation}
Using the definition of $S_i$ from equation (\ref{Selection}) and substituting in equations (\ref{perceivedreturns}) and (\ref{Perceived_Returns_CF}), it follows that
\begin{equation*}
\begin{split}
\mathbb{E}[\xi_i|S_i=1,X_i,Price_i,u_i]
=
\mathbb{E}\big[
&
\xi_i
\big|
\big(
-\xi_i \leq X_i\theta-{Price_i}+u_i\rho
\big)
,X_i,Price_i,u_i\big]
\\
=
-\mathbb{E}\big[-
&
\xi_i
\big|
\big(
-\xi_i \leq X_i\theta-{Price_i}+u_i\rho
\big)
,X_i,Price_i,u_i\big],
\end{split}
\end{equation*}
which allows us to rewrite
\begin{equation}
\mathbb{E}[\xi_i|S_i=1,X_i,Price_i,u_i]
=
\sigma_\xi
\frac{
\phi\big(X_i\theta^*_\xi-{Price_i}\gamma^*_\xi+u_i\rho^*_\xi\big)
}
{
\Phi\big(X_i\theta^*_\xi-{Price_i}\gamma^*_\xi+u_i\rho^*_\xi\big)
}
\end{equation}
using Expression (\ref{error_MI_endo}) and applying the symmetry of the normal distribution. Equation (\ref{IM_rp2}) follows by applying this equality to (\ref{rp2_rearrange}) and dividing each side of the resulting inequality by $\sigma_\xi$. $\blacksquare$

\begin{lem}
Given $\frac{\rho}{1-\rho}\geq 0$, equations $(\ref{Selection})$, $(\ref{perceivedreturns})$, $(\ref{Price_Z})$, and $(\ref{Perceived_Returns_CF})$ imply
\begin{equation}\label{rp2_drop_uv}
\begin{gathered}
\mathbb{E}\Bigg[
S_i
\Bigg(
-
u_i\gamma^*_\xi
+
\frac
{\phi\big(X_i\theta^*_\xi-Price_i\gamma^*_\xi\big)}
{\Phi\big(X_i\theta^*_\xi-Price_i\gamma^*_\xi\big)}
\Bigg)
\Bigg|
Z_i
\Bigg]\\
\geq\\
\mathbb{E}\Bigg[
S_i
\Bigg(
-
u_i(\gamma^*_\xi-\rho^*_\xi)
+
\frac
{\phi\big(X_i\theta^*_\xi-Price_i\gamma^*_\xi+u_i\rho^*_\xi\big)}
{\Phi\big(X_i\theta^*_\xi-Price_i\gamma^*_\xi+u_i\rho^*_\xi\big)}
\Bigg)
\Bigg|
Z_i
\Bigg]
\end{gathered}
\end{equation}
\end{lem}
\noindent \textbf{Proof:} Using the definition of $Price_i$ from equation $(\ref{Price_Z})$, it follows that
\begin{equation}
\begin{gathered}
\mathbb{E}\Bigg[
S_i
\Bigg(
-
u_i(\gamma^*_\xi-\rho^*_\xi)
+
\frac
{\phi\big(X_i\theta^*_\xi-Price_i\gamma^*_\xi+u_i\rho^*_\xi\big)}
{\Phi\big(X_i\theta^*_\xi-Price_i\gamma^*_\xi+u_i\rho^*_\xi\big)}
\Bigg)
\Bigg|
Z_i
\Bigg]\\
=\\
\mathbb{E}\Bigg[
S_i
\Bigg(
-
u_i(\gamma^*_\xi-\rho^*_\xi)
+
\frac
{\phi\big(X_i\theta^*_\xi-Z_i\delta\gamma^*_\xi-u_i(\gamma^*_\xi-\rho^*_\xi)\big)}
{\Phi\big(X_i\theta^*_\xi-Z_i\delta\gamma^*_\xi-u_i(\gamma^*_\xi-\rho^*_\xi)\big)}
\Bigg)
\Bigg|
Z_i
\Bigg].
\end{gathered}
\end{equation}
Law of iterated expectations and $S_i\in \{0,1\}$ implies
\begin{equation}\label{rp2_error_add_S0Z}
\begin{gathered}
\mathbb{E}\Bigg[
S_i
\Bigg(
-
u_i(\gamma^*_\xi-\rho^*_\xi)
+
\frac
{\phi\big(X_i\theta^*_\xi-Z_i\delta\gamma^*_\xi-u_i(\gamma^*_\xi-\rho^*_\xi)\big)}
{\Phi\big(X_i\theta^*_\xi-Z_i\delta\gamma^*_\xi-u_i(\gamma^*_\xi-\rho^*_\xi)\big)}
\Bigg)
\Bigg|
Z_i
\Bigg]\\
=\\
\mathbb{E}\Bigg[
S_i
\mathbb{E}
\bigg[
-
u_i(\gamma^*_\xi-\rho^*_\xi)
+
\frac
{\phi\big(X_i\theta^*_\xi-Z_i\delta\gamma^*_\xi-u_i(\gamma^*_\xi-\rho^*_\xi)\big)}
{\Phi\big(X_i\theta^*_\xi-Z_i\delta\gamma^*_\xi-u_i(\gamma^*_\xi-\rho^*_\xi)\big)}
\Bigg)
\bigg|
S_i=1,Z_i
\bigg]
\Bigg|
Z_i
\Bigg].
\end{gathered}
\end{equation}
Because
\begin{equation*}
\frac{
\partial \frac
{\phi(-x)}
{\Phi(-x)}}
{\partial x}
\in (0,1),
\end{equation*}
it follows that the expression
\begin{equation}
-u_i(\gamma^*_\xi-\rho^*_\xi)
+
\frac
{\phi\big(X_i\theta^*_\xi-Z_i\delta\gamma^*_\xi-u_i(\gamma^*_\xi-\rho^*_\xi)\big)}
{\Phi\big(X_i\theta^*_\xi-Z_i\delta\gamma^*_\xi-u_i(\gamma^*_\xi-\rho^*_\xi)\big)}
\end{equation}
is monotonically decreasing in $u_i(\gamma^*_\xi-\rho^*_\xi)$. It follows then that adding a negative value to this value will increase the value of the function. From (\ref{error_inequality}) and the condition $\frac{\rho^*_\xi}{\gamma^*_\xi-\rho^*_\xi} \geq 0$, $\frac{\rho^*_\xi}{\gamma^*_\xi-\rho^*_\xi}\mathbb{E}[u_i(\gamma^*_\xi-\rho^*_\xi)|S_i=1,Z_i] \leq 0$ $\forall i$, so it follows that
\begin{equation}\label{rp2_add_constantS1}
\begin{gathered}
\mathbb{E}
\bigg
[
-
u_i(\gamma^*_\xi-\rho^*_\xi)-\mathbb{E}[u_i\rho^*_\xi|S_i=1,Z_i]
\\+
\frac
{\phi\big(X_i\theta^*_\xi-Z_i\delta\gamma^*_\xi-u_i(\gamma^*_\xi-\rho^*_\xi)-\mathbb{E}[u_i\rho^*_\xi|S_i=1,Z_i]\big)}
{\Phi\big(X_i\theta^*_\xi-Z_i\delta\gamma^*_\xi-u_i(\gamma^*_\xi-\rho^*_\xi)-\mathbb{E}[u_i\rho^*_\xi|S_i=1,Z_i]\big)}
\bigg
|
S_i=1,Z_i
\bigg
]
\\
=
\\
\mathbb{E}
\bigg
[
-
\big(
u_i(\gamma^*_\xi-\rho^*_\xi)
+
\frac{\rho^*_\xi}{\gamma^*_\xi-\rho^*_\xi}
\mathbb{E}[u_i(\gamma^*_\xi-\rho^*_\xi)|S_i=1,Z_i]
\big)
\\
+
\frac
{\phi\Big(X_i\theta^*_\xi-Z_i\delta\gamma^*_\xi-\big(u_i(\gamma^*_\xi-\rho^*_\xi)+
\frac{\rho^*_\xi}{\gamma^*_\xi-\rho^*_\xi}
\mathbb{E}[u_i(\gamma^*_\xi-\rho^*_\xi)|S_i=1,Z_i]\big)\Big)}
{\Phi\Big(X_i\theta^*_\xi-Z_i\delta\gamma^*_\xi-\big(u_i(\gamma^*_\xi-\rho^*_\xi)+
\frac{\rho^*_\xi}{\gamma^*_\xi-\rho^*_\xi}
\mathbb{E}[u_i(\gamma^*_\xi-\rho^*_\xi)|S_i=1,Z_i]\big)\Big)}
\bigg
|
S_i=1,Z_i
\bigg
]
\\
\geq \\
\mathbb{E}
\bigg
[
-
u_i(\gamma^*_\xi-\rho^*_\xi)
+
\frac
{\phi\big(X_i\theta^*_\xi-Z_i\delta\gamma^*_\xi-u_i(\gamma^*_\xi-\rho^*_\xi)\big)}
{\Phi\big(X_i\theta^*_\xi-Z_i\delta\gamma^*_\xi-u_i(\gamma^*_\xi-\rho^*_\xi)\big)}
\bigg
|
S_i=1,Z_i
\bigg
],
\end{gathered}
\end{equation}
where the second line relates to the third by this addition, and the first relates to the second by algebraic simplifications. Finally, because the term 
\begin{equation*}
    \frac
{\phi\big(X_i\theta^*_\xi-Z_i\delta\gamma^*_\xi-u_i(\gamma^*_\xi-\rho^*_\xi)-\mathbb{E}[u_i\rho^*_\xi|S_i=1,Z_i]\big)}
{\Phi\big(X_i\theta^*_\xi-Z_i\delta\gamma^*_\xi-u_i(\gamma^*_\xi-\rho^*_\xi)-\mathbb{E}[u_i\rho^*_\xi|S_i=1,Z_i]\big)}
\end{equation*}
is globally convex in $-u_i$, the function is globally convex in $-u_i$. It follows that
\begin{equation}
\begin{gathered}
\mathbb{E}
\bigg[-
u_i\gamma^*_\xi+
\frac
{\phi\big(X_i\theta^*_\xi-Z_i\delta\gamma^*_\xi-u_i\gamma^*_\xi\big)}
{\Phi\big(X_i\theta^*_\xi-Z_i\delta\gamma^*_\xi-u_i\gamma^*_\xi\big)}
\bigg
|S_i=1,Z_i
\bigg
]
\\
\geq \\
\mathbb{E}
\bigg[
-
u_i(\gamma^*_\xi-\rho^*_\xi)
-\mathbb{E}[u_i\rho^*_\xi|S_i=1,Z_i]
\\+
\frac
{\phi\big(X_i\theta^*_\xi-Z_i\delta\gamma^*_\xi-u_i(\gamma^*_\xi-\rho^*_\xi)-\mathbb{E}[u_i\rho^*_\xi|S_i=1,Z_i]\big)}
{\Phi\big(X_i\theta^*_\xi-Z_i\delta\gamma^*_\xi-u_i(\gamma^*_\xi-\rho^*_\xi)-\mathbb{E}[u_i\rho^*_\xi|S_i=1,Z_i]\big)}
\bigg
|S_i=1,Z_i
\bigg
]
\end{gathered}
\end{equation}
by Jensen's inequality. Combining this inequality with that in (\ref{rp2_add_constantS1}) yields the result 
\begin{equation}
\begin{gathered}
\mathbb{E}
\bigg[
-
u_i\gamma^*_\xi+
\frac
{\phi\big(X_i\theta^*_\xi-Z_i\delta\gamma^*_\xi-u_i\gamma^*_\xi\big)}
{\Phi\big(X_i\theta^*_\xi-Z_i\delta\gamma^*_\xi-u_i\gamma^*_\xi\big)}
\bigg
|S_i=1,Z_i
\bigg
]
\\
\geq \\
\mathbb{E}
\bigg
[
-
u_i(\gamma^*_\xi-\rho^*_\xi)
+
\frac
{\phi\big(X_i\theta^*_\xi-Z_i\delta\gamma^*_\xi-u_i(\gamma^*_\xi-\rho^*_\xi)\big)}
{\Phi\big(X_i\theta^*_\xi-Z_i\delta\gamma^*_\xi-u_i(\gamma^*_\xi-\rho^*_\xi)\big)}
\bigg
|
S_i=1,Z_i
\bigg
].
\end{gathered}
\end{equation}
It follows immediately that 
\begin{equation}
\begin{gathered}
\mathbb{E}
\Bigg[
S_i
\mathbb{E}
\bigg[
-
u_i\gamma^*_\xi+
\frac
{\phi\big(X_i\theta^*_\xi-Z_i\delta\gamma^*_\xi-u_i\gamma^*_\xi\big)}
{\Phi\big(X_i\theta^*_\xi-Z_i\delta\gamma^*_\xi-u_i\gamma^*_\xi\big)}
\bigg
|S_i=1,Z_i
\bigg
]
\Bigg
|Z_i
\Bigg
]
\\
\geq \\
\mathbb{E}
\Bigg
[
S_i
\mathbb{E}
\bigg[
-
u_i(\gamma^*_\xi-\rho^*_\xi)
+
\frac
{\phi\big(X_i\theta^*_\xi-Z_i\delta\gamma^*_\xi-u_i(\gamma^*_\xi-\rho^*_\xi)\big)}
{\Phi\big(X_i\theta^*_\xi-Z_i\delta\gamma^*_\xi-u_i(\gamma^*_\xi-\rho^*_\xi)\big)}
\Bigg
|
S_i=1,Z_i
\bigg
]
\Bigg
|Z_i
\Bigg
].
\end{gathered}
\end{equation}
Equation (\ref{rp2_drop_uv}) follows from this by substituting the definition of $Price_i$ from (\ref{Price_Z}) and applying the law of iterated expectations. $\blacksquare$

\noindent\textbf{Corollary 2} \textit{Given (\ref{IM_rp2}), (\ref{rp2_drop_uv}), and the definition of $Price_i$ given in $(\ref{Price_Z})$, it follows that}
\begin{equation}\label{cor1_2}
\mathbb{E}\Bigg[
-(1-S_i) (X_i\theta^*_\xi-Price_i\gamma^*_\xi)
+
S_i
\frac
{\phi(X_i\theta^*_\xi-Price_i\gamma^*_\xi)}
{\Phi(X_i\theta^*_\xi-Price_i\gamma^*_\xi)}
\Bigg|Z_i\Bigg]
\geq 0.
\end{equation}

\noindent\textbf{Proof:} The result follows from equations (\ref{IM_rp2}), (\ref{rp2_drop_uv}), that $\mathbb{E}[S_iu_i=-(1-S_i)u_i|Z_i]$, and the definition of $Price_i$ given in $(\ref{Price_Z})$. $\blacksquare$

\noindent \textbf{Proof of Robustness of Revealed Preference Inequalities to Endogeneity:} Combining equations (\ref{cor1_1}) and (\ref{cor1_2}) provides both inequalities defined in equation (\ref{RPMI}). $\blacksquare$

\subsection{Argument for Odds-Based Inequality Robustness to Endogeneity}
The following argument is constructed as a proof, where the components of the argument that do not meet the standards of a proof are discussed as they arise.
\begin{lem}
Equations $(\ref{Selection})$, $(\ref{perceivedreturns})$, $(\ref{Perceived_Returns_CF})$, $(\ref{error_MI_endo})$, and the assumption that the distribution of $Z_i$ is degenerate conditional on $(X_i,Price_i,u_i)$ imply that
\begin{equation}\label{ob1_rho}
\mathbb{E}
\Bigg[
S_i
\frac
{
1-\Phi
\big
(
X_i\theta^*_\xi-Price_i\gamma^*_\xi+u_i\rho^*_\xi
\big
)
}
{
\Phi
\big
(
X_i\theta^*_\xi-Price_i\gamma^*_\xi+u_i\rho^*_\xi
\big
)
}
-(1-S_i)
\Bigg
|
Z_i
\Bigg
]
\geq 
0
\end{equation}.
\end{lem}
\noindent \textbf{Proof:} Expression (\ref{error_MI_endo}) implies that 
\begin{equation*}
S_i
-
\mathbbm{1}\{X_i\theta-Price_i+u_i\rho+\xi_i
\geq
0
\}
\geq
0,
\end{equation*}
or, equivalently,
\begin{equation*}
\begin{split}
1
-
\mathbbm{1}\{X_i\theta-Price_i+u_i\rho+\xi_i
\geq
0
\}
-
(1-S_i)
&\geq
0,\\
\mathbbm{1}\{X_i\theta-Price_i+u_i\rho+\xi_i
\leq
0
\}
-
(1-S_i)
&\geq
0,
\end{split}
\end{equation*}
for all $i$. Given that this inequality holds for all individuals, it will also hold in expectation, conditional on any set of variables, across individuals. It follows that
\begin{equation*}
\mathbb{E}[\mathbbm{1}\{X_i\theta-Price_i+u_i\rho+\xi_i
\leq
0
\}
-
(1-S_i)
|
X_i,Price_i,u_i
]
\geq
0.
\end{equation*}
The distributional assumption in (\ref{error_MI_endo}) implies
\begin{equation*}
\mathbb{E}[
1-\Phi(
X_i\theta^*_\xi-Price_i\gamma^*_\xi+u_i\rho^*_\xi
)
-
(1-S_i)
|
X_i,Price_i,u_i
]
\geq
0.
\end{equation*}
Dividing through by $\Phi(
X_i\theta^*_\xi-Price_i\gamma^*_\xi+u_i\rho^*_\xi)$ yields
\begin{equation*}
\begin{split}
\mathbb{E}
\bigg
[
&\frac{
1-\Phi(
X_i\theta^*_\xi-Price_i\gamma^*_\xi+u_i\rho^*_\xi
)
}
{
\Phi(
X_i\theta^*_\xi-Price_i\gamma^*_\xi+u_i\rho^*_\xi
)
}\\
&\mbox{\hspace{15mm}}
-
\frac{
(1-S_i)
}
{
\Phi(
X_i\theta^*_\xi-Price_i\gamma^*_\xi+u_i\rho^*_\xi
)
}
\bigg
|
X_i,Price_i,u_i
\bigg
]
\geq
0.
\end{split}
\end{equation*}
Adding and subtracting $1-S_i$ gives
\begin{equation*}
\begin{split}
\mathbb{E}
\bigg
[
&\frac{
1-\Phi(
X_i\theta^*_\xi-Price_i\gamma^*_\xi+u_i\rho^*_\xi
)
}
{
\Phi(
X_i\theta^*_\xi-Price_i\gamma^*_\xi+u_i\rho^*_\xi
)
}\\
&\mbox{\hspace{10mm}}
-
\bigg
(
1-1
+
\frac{
1
}
{
\Phi(
X_i\theta^*_\xi-Price_i\gamma^*_\xi+u_i\rho^*_\xi
)
}
\bigg
)
(1-S_i)
\bigg
|
X_i,Price_i,u_i
\bigg
]
\geq
0,
\end{split}
\end{equation*}
which we can rearrange into
\begin{equation*}
\begin{split}
\mathbb{E}
\bigg
[
&\frac{
1-\Phi(
X_i\theta^*_\xi-Price_i\gamma^*_\xi+u_i\rho^*_\xi
)
}
{
\Phi(
X_i\theta^*_\xi-Price_i\gamma^*_\xi+u_i\rho^*_\xi
)
}\\
&\mbox{\hspace{10mm}}
-
\bigg
(
1
+
\frac{
1-\Phi(
X_i\theta^*_\xi-Price_i\gamma^*_\xi+u_i\rho^*_\xi
)
}
{
\Phi(
X_i\theta^*_\xi-Price_i\gamma^*_\xi+u_i\rho^*_\xi
)
}
\bigg
)
(1-S_i)
\bigg
|
X_i,Price_i,u_i
\bigg
]
\geq
0,
\end{split}
\end{equation*}
which can then be rearranged into
\begin{equation*}
\mathbb{E}
\bigg
[
S_i
\frac{
1-\Phi(
X_i\theta^*_\xi-Price_i\gamma^*_\xi+u_i\rho^*_\xi
)
}
{
\Phi(
X_i\theta^*_\xi-Price_i\gamma^*_\xi+u_i\rho^*_\xi
)
}
-
(
1-S_i
)
\bigg
|
X_i,Price_i,u_i
\bigg
]
\geq
0.
\end{equation*}
Equation (\ref{ob1_rho}) follows from the law of iterated expectations and the assumption that the distribution of $Z_i$ conditional on $(X_i,Price_i,u_i)$ is degenerate. $\blacksquare$

\begin{lem}
If equations $(\ref{Selection})$, $(\ref{perceivedreturns})$, $(\ref{Price_Z})$, and $(\ref{Perceived_Returns_CF})$ hold and $\frac{\rho}{1-\rho}\geq0$, then 
\begin{equation}\label{ob1_addEuv}
\begin{gathered}
\mathbb{E}
\Bigg[
S_i
\frac
{
1-\Phi
\big
(
X_i\theta^*_\xi-Price_i\gamma^*_\xi
\big
)
}
{
\Phi
\big
(
X_i\theta^*_\xi-Price_i\gamma^*_\xi
\big
)
}
\Bigg
|
Z_i
\Bigg
]\\
\geq\\
\mathbb{E}
\Bigg[
S_i
\frac
{
1-\Phi
\big
(
X_i\theta^*_\xi-Price_i\gamma^*_\xi+u_i\rho^*_\xi
\big
)
}
{
\Phi
\big
(
X_i\theta^*_\xi-Price_i\gamma^*_\xi+u_i\rho^*_\xi
\big
)
}
\Bigg
|
Z_i
\Bigg
].
\end{gathered}
\end{equation}
\end{lem}
\noindent \textbf{Argument: } Substituting the definition of $Price_i$ from equation (\ref{Price_Z}), we have that
\begin{equation*}
\begin{gathered}
\mathbb{E}
\Bigg[
S_i
\frac
{
1-\Phi
\big
(
X_i\theta^*_\xi-Z_i\delta\gamma^*_\xi-u_i(\gamma^*_\xi-\rho^*_\xi)
\big
)
}
{
\Phi
\big
(
X_i\theta^*_\xi-Z_i\delta\gamma^*_\xi-u_i(\gamma^*_\xi-\rho^*_\xi)
\big
)
}
\Bigg
|
Z_i
\Bigg
]\\
=\\
\mathbb{E}
\Bigg[
S_i
\frac
{
1-\Phi
\big
(
X_i\theta^*_\xi-Price_i\gamma^*_\xi+u_i\rho^*_\xi
\big
)
}
{
\Phi
\big
(
X_i\theta^*_\xi-Price_i\gamma^*_\xi+u_i\rho^*_\xi
\big
)
}
\Bigg
|
Z_i
\Bigg
].
\end{gathered}
\end{equation*}
Because $S_i \in \{0,1\}$, it follows that 
\begin{equation*}
\begin{gathered}
\mathbb{E}
\Bigg[
S_i
\frac
{
1-\Phi
\big
(
X_i\theta^*_\xi-Z_i\delta\gamma^*_\xi-u_i(\gamma^*_\xi-\rho^*_\xi)
\big
)
}
{
\Phi
\big
(
X_i\theta^*_\xi-Z_i\delta\gamma^*_\xi-u_i(\gamma^*_\xi-\rho^*_\xi)
\big
)
}
\Bigg
|
Z_i
\Bigg
]
 \geq 0 
 \\ 
 \iff
  \\
\mathbb{E}
\Bigg[
\frac
{
1-\Phi
\big
(
X_i\theta^*_\xi-Z_i\delta\gamma^*_\xi-u_i(\gamma^*_\xi-\rho^*_\xi)
\big
)
}
{
\Phi
\big
(
X_i\theta^*_\xi-Z_i\delta\gamma^*_\xi-u_i(\gamma^*_\xi-\rho^*_\xi)
\big
)
}
\Bigg
|
S_i=1, Z_i
\Bigg
]
 \geq 0.
\end{gathered}
\end{equation*}
Proving the argument requires that 
\begin{equation*}
\begin{gathered}
\mathbb{E}
\Bigg[
\frac
{
1-\Phi
\big
(
X_i\theta^*_\xi-Z_i\delta\gamma^*_\xi-u_i(\gamma^*_\xi-\rho^*_\xi)-u_i\rho^*_\xi
\big
)
}
{
\Phi
\big
(
X_i\theta^*_\xi-Z_i\delta\gamma^*_\xi-u_i(\gamma^*_\xi-\rho^*_\xi)-u_i\rho^*_\xi
\big
)
}
\Bigg
|
S_i=1, Z_i
\Bigg
] \\
 \geq \\
 \mathbb{E}
\Bigg[
\frac
{
1-\Phi
\big
(
X_i\theta^*_\xi-Z_i\delta\gamma^*_\xi-u_i(\gamma^*_\xi-\rho^*_\xi)
\big
)
}
{
\Phi
\big
(
X_i\theta^*_\xi-Z_i\delta\gamma^*_\xi-u_i(\gamma^*_\xi-\rho^*_\xi)
\big
)
}
\Bigg
|
S_i=1, Z_i
\Bigg
].
\end{gathered}
\end{equation*}
Given that \hfill
\begin{equation*}
    \frac
{
1-\Phi
(
x)
}
{
\Phi
(
x)
}
\end{equation*}
is globally convex in $x$, Jensen's inequality implies that
\begin{equation*}
\begin{gathered}
\mathbb{E}
\Bigg[
\frac
{
1-\Phi
\big
(
X_i\theta^*_\xi-Z_i\delta\gamma^*_\xi-u_i(\gamma^*_\xi-\rho^*_\xi)-u_i\rho^*_\xi
\big
)
}
{
\Phi
\big
(
X_i\theta^*_\xi-Z_i\delta\gamma^*_\xi-u_i(\gamma^*_\xi-\rho^*_\xi)-u_i\rho^*_\xi
\big
)
}
\Bigg
|
S_i=1, Z_i
\Bigg
] \\
 \geq \\
\mathbb{E}
\Bigg[
\frac
{
1-\Phi
\big
(
X_i\theta^*_\xi-Z_i\delta\gamma^*_\xi-u_i(\gamma^*_\xi-\rho^*_\xi)-\mathbb{E}[u_i\rho^*_\xi|S_i=1,Z_i]
\big
)
}
{
\Phi
\big
(
X_i\theta^*_\xi-Z_i\delta\gamma^*_\xi-u_i(\gamma^*_\xi-\rho^*_\xi)-\mathbb{E}[u_i\rho^*_\xi|S_i=1,Z_i]
\big
)
}
\Bigg
|
S_i=1, Z_i
\Bigg
].
\end{gathered}
\end{equation*}
Meanwhile, (\ref{error_inequality}) implies
\begin{equation*}
\begin{gathered}
\mathbb{E}
\Bigg[
\frac
{
1-\Phi
\big
(
X_i\theta^*_\xi-Z_i\delta\gamma^*_\xi-u_i(\gamma^*_\xi-\rho^*_\xi)-\mathbb{E}[u_i\rho^*_\xi|S_i=1,Z_i]
\big
)
}
{
\Phi
\big
(
X_i\theta^*_\xi-Z_i\delta\gamma^*_\xi-u_i(\gamma^*_\xi-\rho^*_\xi)-\mathbb{E}[u_i\rho^*_\xi|S_i=1,Z_i]
\big
)
}
\Bigg
|
S_i=1, Z_i
\Bigg
] \\
 \leq \\
\mathbb{E}
\Bigg[
\frac
{
1-\Phi
\big
(
X_i\theta^*_\xi-Z_i\delta\gamma^*_\xi-u_i(\gamma^*_\xi-\rho^*_\xi)
\big
)
}
{
\Phi
\big
(
X_i\theta^*_\xi-Z_i\delta\gamma^*_\xi-u_i(\gamma^*_\xi-\rho^*_\xi)
\big
)
}
\Bigg
|
S_i=1, Z_i
\Bigg
].
\end{gathered}
\end{equation*}
Combining these inequalities yields
\begin{equation*}
\begin{gathered}
\mathbb{E}
\Bigg[
\frac
{
1-\Phi
\big
(
X_i\theta^*_\xi-Z_i\delta\gamma^*_\xi-u_i(\gamma^*_\xi-\rho^*_\xi)-u_i\rho^*_\xi
\big
)
}
{
\Phi
\big
(
X_i\theta^*_\xi-Z_i\delta\gamma^*_\xi-u_i(\gamma^*_\xi-\rho^*_\xi)-u_i\rho^*_\xi
\big
)
}
\Bigg
|
S_i=1, Z_i
\Bigg
] \\
 \geq \\
\mathbb{E}
\Bigg[
\frac
{
1-\Phi
\big
(
X_i\theta^*_\xi-Z_i\delta\gamma^*_\xi-u_i(\gamma^*_\xi-\rho^*_\xi)-\mathbb{E}[u_i\rho^*_\xi|S_i=1,Z_i]
\big
)
}
{
\Phi
\big
(
X_i\theta^*_\xi-Z_i\delta\gamma^*_\xi-u_i(\gamma^*_\xi-\rho^*_\xi)-\mathbb{E}[u_i\rho^*_\xi|S_i=1,Z_i]
\big
)
}
\Bigg
|
S_i=1, Z_i
\Bigg
] \\
 \leq \\
\mathbb{E}
\Bigg[
\frac
{
1-\Phi
\big
(
X_i\theta^*_\xi-Z_i\delta\gamma^*_\xi-u_i(\gamma^*_\xi-\rho^*_\xi)
\big
)
}
{
\Phi
\big
(
X_i\theta^*_\xi-Z_i\delta\gamma^*_\xi-u_i(\gamma^*_\xi-\rho^*_\xi)
\big
)
}
\Bigg
|
S_i=1, Z_i
\Bigg
].
\end{gathered}
\end{equation*}
Thus, the argument holds if the first inequality dominates the second. There is good reason to believe that this will be the case. The first inequality arises from $Var(u_i|S_i=1, Z_i)$ (through Jensen's inequality), while the second arises from $\mathbb{E}[u_i|S_i=1,Z_i]$. Three points are salient here. First, given that $\mathbb{E}[u_i|Z_i] = 0$, the value of $\mathbb{E}[u_i|S_i=1,Z_i]$ is a monotonic function of $Var(u_i|Z_i)$. Second, the function $(1-\Phi(x))/\Phi(x)$ has a very large second derivative for most of its support, such that the application of Jensen's inequality will have a large effect on the inequality. Thirdly, because $\rho^*_\xi$ is constrained to be small relative to $\gamma^*_\xi$, $\mathbb{E}[u_i|S_i=1,Z_i]$ is likely to have a relatively small effect on the inequality.

Equation (\ref{ob1_addEuv}) follows from the preceding inequalities if the first inequality dominates the second by performing simple algebraic manipulations and applying the definition of $Price_i$ given in (\ref{Price_Z}). $\blacksquare$

\begin{lem}
Equations $(\ref{Selection})$, $(\ref{perceivedreturns})$, $(\ref{Perceived_Returns_CF})$, $(\ref{error_MI_endo})$, and the assumption that the distribution of $Z_i$ is degenerate conditional on $(X_i,Price_i,u_i)$ imply that
\begin{equation}\label{ob2_rho}
\mathbb{E}
\Bigg[
(1-S_i)
\frac
{
\Phi
\big
(
X_i\theta^*_\xi-Price_i\gamma^*_\xi+u_i\rho^*_\xi
\big
)
}
{
1-\Phi
\big
(
X_i\theta^*_\xi-Price_i\gamma^*_\xi+u_i\rho^*_\xi
\big
)
}
-S_i
\Bigg
|
Z_i
\Bigg
]
\geq 
0
\end{equation}.
\end{lem}
\noindent \textbf{Proof:} Expression (\ref{error_MI_endo}) implies that 
\begin{equation*}
\mathbbm{1}\{X_i\theta-Price_i+u_i\rho+\xi_i
\geq
0
\}
-S_i
\geq
0,
\end{equation*}
Given that this inequality holds for all individuals, it will also hold in expectation, conditional on any set of variables, across individuals. It follows that
\begin{equation*}
\mathbb{E}[\mathbbm{1}\{X_i\theta-Price_i+u_i\rho+\xi_i
\leq
0
\}
-
S_i
|
X_i,Price_i,u_i
]
\geq
0.
\end{equation*}
The distributional assumption in (\ref{error_MI_endo}) implies
\begin{equation*}
\mathbb{E}[
\Phi(
X_i\theta^*_\xi-Price_i\gamma^*_\xi+u_i\rho^*_\xi
)
-
S_i
|
X_i,Price_i,u_i
]
\geq
0.
\end{equation*}
Dividing through by $1-\Phi(
X_i\theta^*_\xi-Price_i\gamma^*_\xi+u_i\rho^*_\xi)$ yields
\begin{equation*}
\begin{split}
\mathbb{E}
\bigg
[
&\frac{
\Phi(
X_i\theta^*_\xi-Price_i\gamma^*_\xi+u_i\rho^*_\xi
)
}
{
1-\Phi(
X_i\theta^*_\xi-Price_i\gamma^*_\xi+u_i\rho^*_\xi
)
}\\
&\mbox{\hspace{15mm}}
-
\frac{
S_i
}
{
1-\Phi(
X_i\theta^*_\xi-Price_i\gamma^*_\xi+u_i\rho^*_\xi
)
}
\bigg
|
X_i,Price_i,u_i
\bigg
]
\geq
0.
\end{split}
\end{equation*}
Adding and subtracting $S_i$ gives
\begin{equation*}
\begin{split}
\mathbb{E}
\bigg
[
&\frac{
\Phi(
X_i\theta^*_\xi-Price_i\gamma^*_\xi+u_i\rho^*_\xi
)
}
{
1-\Phi(
X_i\theta^*_\xi-Price_i\gamma^*_\xi+u_i\rho^*_\xi
)
}\\
&\mbox{\hspace{10mm}}
-
\bigg
(
1-1
+
\frac{
1
}
{
1-\Phi(
X_i\theta^*_\xi-Price_i\gamma^*_\xi+u_i\rho^*_\xi
)
}
\bigg
)
S_i
\bigg
|
X_i,Price_i,u_i
\bigg
]
\geq
0,
\end{split}
\end{equation*}
which we can rearrange into
\begin{equation*}
\begin{split}
\mathbb{E}
\bigg
[
&\frac{
\Phi(
X_i\theta^*_\xi-Price_i\gamma^*_\xi+u_i\rho^*_\xi
)
}
{
1-\Phi(
X_i\theta^*_\xi-Price_i\gamma^*_\xi+u_i\rho^*_\xi
)
}\\
&\mbox{\hspace{10mm}}
-
\bigg
(
1
+
\frac{
\Phi(
X_i\theta^*_\xi-Price_i\gamma^*_\xi+u_i\rho^*_\xi
)
}
{
1-\Phi(
X_i\theta^*_\xi-Price_i\gamma^*_\xi+u_i\rho^*_\xi
)
}
\bigg
)
S_i
\bigg
|
X_i,Price_i,u_i
\bigg
]
\geq
0,
\end{split}
\end{equation*}
which is straightforward to rearrange into
\begin{equation*}
\mathbb{E}
\bigg
[
(1-S_i)
\frac{
\Phi(
X_i\theta^*_\xi-Price_i\gamma^*_\xi+u_i\rho^*_\xi
)
}
{
1-\Phi(
X_i\theta^*_\xi-Price_i\gamma^*_\xi+u_i\rho^*_\xi
)
}
-
S_i
\bigg
|
X_i,Price_i,u_i
\bigg
]
\geq
0.
\end{equation*}
Equation (\ref{ob2_rho}) follows from the law of iterated expectations and the assumption that the distribution of $Z_i$ conditional on $(X_i,Price_i,u_i)$ is degenerate. $\blacksquare$

\begin{lem}
If equations $(\ref{Selection})$, $(\ref{perceivedreturns})$, $(\ref{Price_Z})$, and $(\ref{Perceived_Returns_CF})$ hold and $\frac{\rho}{1-\rho}\geq0$, then 
\begin{equation}\label{ob2_addEuv}
\begin{gathered}
\mathbb{E}
\Bigg[
(1-S_i)
\frac
{
\Phi
\big
(
X_i\theta^*_\xi-Price_i\gamma^*_\xi
\big
)
}
{
1-\Phi
\big
(
X_i\theta^*_\xi-Price_i\gamma^*_\xi
\big
)
}
\Bigg
|
Z_i
\Bigg
]\\
\geq\\
\mathbb{E}
\Bigg[
(1-S_i)
\frac
{
\Phi
\big
(
X_i\theta^*_\xi-Price_i\gamma^*_\xi+u_i\rho^*_\xi
\big
)
}
{
1-\Phi
\big
(
X_i\theta^*_\xi-Price_i\gamma^*_\xi+u_i\rho^*_\xi
\big
)
}
\Bigg
|
Z_i
\Bigg
].
\end{gathered}
\end{equation}
\end{lem}
\noindent \textbf{Argument: } Substituting the definition of $Price_i$ from equation (\ref{Price_Z}), we have that
\begin{equation*}
\begin{gathered}
\mathbb{E}
\Bigg[
(1-S_i)
\frac
{
\Phi
\big
(
X_i\theta^*_\xi-Z_i\delta\gamma^*_\xi-u_i(\gamma^*_\xi-\rho^*_\xi)
\big
)
}
{
1-\Phi
\big
(
X_i\theta^*_\xi-Z_i\delta\gamma^*_\xi-u_i(\gamma^*_\xi-\rho^*_\xi)
\big
)
}
\Bigg
|
Z_i
\Bigg
]\\
=\\
\mathbb{E}
\Bigg[
(1-S_i)
\frac
{
\Phi
\big
(
X_i\theta^*_\xi-Price_i\gamma^*_\xi+u_i\rho^*_\xi
\big
)
}
{
1-\Phi
\big
(
X_i\theta^*_\xi-Price_i\gamma^*_\xi+u_i\rho^*_\xi
\big
)
}
\Bigg
|
Z_i
\Bigg
].
\end{gathered}
\end{equation*}
Because $S_i \in \{0,1\}$, it follows that 
\begin{equation*}
\begin{gathered}
\mathbb{E}
\Bigg[
(1-S_i)
\frac
{
\Phi
\big
(
X_i\theta^*_\xi-Z_i\delta\gamma^*_\xi-u_i(\gamma^*_\xi-\rho^*_\xi)
\big
)
}
{
1-\Phi
\big
(
X_i\theta^*_\xi-Z_i\delta\gamma^*_\xi-u_i(\gamma^*_\xi-\rho^*_\xi)
\big
)
}
\Bigg
|
Z_i
\Bigg
]
 \geq 0 
 \\ 
 \iff
  \\
\mathbb{E}
\Bigg[
\frac
{
\Phi
\big
(
X_i\theta^*_\xi-Z_i\delta\gamma^*_\xi-u_i(\gamma^*_\xi-\rho^*_\xi)
\big
)
}
{
1-\Phi
\big
(
X_i\theta^*_\xi-Z_i\delta\gamma^*_\xi-u_i(\gamma^*_\xi-\rho^*_\xi)
\big
)
}
\Bigg
|
S_i=0, Z_i
\Bigg
]
 \geq 0.
\end{gathered}
\end{equation*}
Proving the argument requires that 
\begin{equation*}
\begin{gathered}
\mathbb{E}
\Bigg[
\frac
{
\Phi
\big
(
X_i\theta^*_\xi-Z_i\delta\gamma^*_\xi-u_i(\gamma^*_\xi-\rho^*_\xi)-u_i\rho^*_\xi
\big
)
}
{
1-\Phi
\big
(
X_i\theta^*_\xi-Z_i\delta\gamma^*_\xi-u_i(\gamma^*_\xi-\rho^*_\xi)-u_i\rho^*_\xi
\big
)
}
\Bigg
|
S_i=0, Z_i
\Bigg
] \\
 \geq \\
 \mathbb{E}
\Bigg[
\frac
{
\Phi
\big
(
X_i\theta^*_\xi-Z_i\delta\gamma^*_\xi-u_i(\gamma^*_\xi-\rho^*_\xi)
\big
)
}
{
1-\Phi
\big
(
X_i\theta^*_\xi-Z_i\delta\gamma^*_\xi-u_i(\gamma^*_\xi-\rho^*_\xi)
\big
)
}
\Bigg
|
S_i=0, Z_i
\Bigg
].
\end{gathered}
\end{equation*}
Given that \hfill
\begin{equation*}
    \frac
{
\Phi
(
x)
}
{
1-\Phi
(
x)
}
\end{equation*}
is globally convex in $x$, Jensen's inequality implies that
\begin{equation*}
\begin{gathered}
\mathbb{E}
\Bigg[
\frac
{
\Phi
\big
(
X_i\theta^*_\xi-Z_i\delta\gamma^*_\xi-u_i(\gamma^*_\xi-\rho^*_\xi)-u_i\rho^*_\xi
\big
)
}
{
1-\Phi
\big
(
X_i\theta^*_\xi-Z_i\delta\gamma^*_\xi-u_i(\gamma^*_\xi-\rho^*_\xi)-u_i\rho^*_\xi
\big
)
}
\Bigg
|
S_i=0, Z_i
\Bigg
] \\
 \geq \\
\mathbb{E}
\Bigg[
\frac
{
\Phi
\big
(
X_i\theta^*_\xi-Z_i\delta\gamma^*_\xi-u_i(\gamma^*_\xi-\rho^*_\xi)-\mathbb{E}[u_i\rho^*_\xi|S_i=0,Z_i]
\big
)
}
{
1-\Phi
\big
(
X_i\theta^*_\xi-Z_i\delta\gamma^*_\xi-u_i(\gamma^*_\xi-\rho^*_\xi)-\mathbb{E}[u_i\rho^*_\xi|S_i=0,Z_i]
\big
)
}
\Bigg
|
S_i=0, Z_i
\Bigg
].
\end{gathered}
\end{equation*}
Meanwhile, (\ref{error_inequality}) implies
\begin{equation*}
\begin{gathered}
\mathbb{E}
\Bigg[
\frac
{
\Phi
\big
(
X_i\theta^*_\xi-Z_i\delta\gamma^*_\xi-u_i(\gamma^*_\xi-\rho^*_\xi)-\mathbb{E}[u_i\rho^*_\xi|S_i=0,Z_i]
\big
)
}
{
1-\Phi
\big
(
X_i\theta^*_\xi-Z_i\delta\gamma^*_\xi-u_i(\gamma^*_\xi-\rho^*_\xi)-\mathbb{E}[u_i\rho^*_\xi|S_i=0,Z_i]
\big
)
}
\Bigg
|
S_i=0, Z_i
\Bigg
] \\
 \leq \\
\mathbb{E}
\Bigg[
\frac
{
\Phi
\big
(
X_i\theta^*_\xi-Z_i\delta\gamma^*_\xi-u_i(\gamma^*_\xi-\rho^*_\xi)
\big
)
}
{
1-\Phi
\big
(
X_i\theta^*_\xi-Z_i\delta\gamma^*_\xi-u_i(\gamma^*_\xi-\rho^*_\xi)
\big
)
}
\Bigg
|
S_i=0, Z_i
\Bigg
].
\end{gathered}
\end{equation*}
Combining these inequalities yields
\begin{equation*}
\begin{gathered}
\mathbb{E}
\Bigg[
\frac
{
\Phi
\big
(
X_i\theta^*_\xi-Z_i\delta\gamma^*_\xi-u_i(\gamma^*_\xi-\rho^*_\xi)-u_i\rho^*_\xi
\big
)
}
{
1-\Phi
\big
(
X_i\theta^*_\xi-Z_i\delta\gamma^*_\xi-u_i(\gamma^*_\xi-\rho^*_\xi)-u_i\rho^*_\xi
\big
)
}
\Bigg
|
S_i=0, Z_i
\Bigg
] \\
 \geq \\
\mathbb{E}
\Bigg[
\frac
{
\Phi
\big
(
X_i\theta^*_\xi-Z_i\delta\gamma^*_\xi-u_i(\gamma^*_\xi-\rho^*_\xi)-\mathbb{E}[u_i\rho^*_\xi|S_i=0,Z_i]
\big
)
}
{
1-\Phi
\big
(
X_i\theta^*_\xi-Z_i\delta\gamma^*_\xi-u_i(\gamma^*_\xi-\rho^*_\xi)-\mathbb{E}[u_i\rho^*_\xi|S_i=0,Z_i]
\big
)
}
\Bigg
|
S_i=0, Z_i
\Bigg
] \\
 \leq \\
\mathbb{E}
\Bigg[
\frac
{
\Phi
\big
(
X_i\theta^*_\xi-Z_i\delta\gamma^*_\xi-u_i(\gamma^*_\xi-\rho^*_\xi)
\big
)
}
{
1-\Phi
\big
(
X_i\theta^*_\xi-Z_i\delta\gamma^*_\xi-u_i(\gamma^*_\xi-\rho^*_\xi)
\big
)
}
\Bigg
|
S_i=0, Z_i
\Bigg
].
\end{gathered}
\end{equation*}
Thus, the argument holds if the first inequality dominates the second. There is good reason to believe that this will be the case. The first inequality arises from $Var(u_i|S_i=0, Z_i)$ (through Jensen's inequality), while the second arises from $\mathbb{E}[u_i|S_i=0,Z_i]$. Three points are salient here. First, given that $\mathbb{E}[u_i|Z_i] = 0$, the value of $\mathbb{E}[u_i|S_i=0,Z_i]$ is a monotonic function of $Var(u_i|Z_i)$. Second, the function $\Phi(x)/(1-\Phi(x))$ has a very large second derivative for most of its support, such that the application of Jensen's inequality will have a large effect on the inequality. Thirdly, because $\rho^*_\xi$ is constrained to be small relative to $\gamma^*_\xi$, $\mathbb{E}[u_i|S_i=0,Z_i]$ is likely to have a relatively small effect on the inequality.

Equation (\ref{ob2_addEuv}) follows from the immediately preceding inequalities if the first inequality dominates the second by performing simple algebraic manipulations and applying the definition of $Price_i$ given in (\ref{Price_Z}). $\blacksquare$

\noindent \textbf{Argument for Odds Based Inequality Robustness to Endogeneity:} Substituting equation (\ref{ob1_addEuv}) into (\ref{ob1_rho}) and equation (\ref{ob2_addEuv}) into (\ref{ob2_rho}) provides the inequalities defined in equation (\ref{OBMI}).
\section{Additional Simulations}\label{appendix:More_Sims}

This section presents additional simulations that include estimated bound on parameters using the moment inequality method described in Appendix \ref{appendix:Moment_Inequalities}. I present a series of variations on the setting described in Section \ref{CF_method}, where the magnitudes and directions of selection and misperception biases vary. These simulations demonstrate the robustness of the control function method to a wide variety of empirical settings, while also demonstrating the performance of the moment inequalities in settings other than that described in Section \ref{MI_method}. I also present simulations with additional explanatory variables in order to demonstrate the computational performance of the different estimators.

As in the body of the paper,
I use the following DGP,
\begin{equation}
\begin{gathered}
Y_i = X_i\beta-\widetilde{Price_i}{}+\epsilon_i \\
\widetilde{Price}_i = {Price}_i+\nu_i \\
Price_i = Z_i\delta +u_i, \\
\end{gathered}
\end{equation}
where for these simulations $Z_i$ is always uncorrelated with $\epsilon_i$, $\nu_i$, and $u_i$, and the nature of the covariance structure on these error terms will determine which methods will and will not provide consistent estimates of perceived returns.
Because perceived prices only differ from realized prices in idiosyncratic ways, $\beta=\theta$ in all the following DGPs.
Finally, I note that the probit will estimate $(\theta,\sigma)$, the control function method will estimate $(\theta, \rho, \sigma_\zeta)$, and the moment inequalities will bound $(\theta,\sigma_\epsilon)$ under the assumptions in Appendix \ref{appendix:Moment_Inequalities} or $(\theta,\sigma_\xi)$ under the assumptions in Appendix \ref{appendix:MI_Proofs}, where these are defined in sections \ref{MLE_method}, \ref{CF_method}, and Appendix \ref{appendix:Moment_Inequalities}.

Each DGP is comprised of $N=10,000$ observations of agents whose decisions are governed by their perceived returns to selection.
I construct the instrument vector as $Z_i = [X_i \mbox{ } z_i]$ where $X_i$ always includes only a constant unless otherwise stated, and $z_i$ is a single known and exogenous instrument. Finally, I assume the constant $\beta_0=1$ and $\delta=[0 \mbox{ } 1]'$ for all DGPs.

\subsection{Known, Exogenous Prices}
I begin with a well-behaved benchmark DGP that corresponds to the setting described in Section \ref{MLE_method}.
I generate data according to
\begin{equation}
\begin{bmatrix}
z_i \\ 
u_i \\
\nu_i \\
\epsilon_i
\end{bmatrix}
\sim 
\mathcal{N}(\textbf{0},\Sigma);
\quad \Sigma = 
\begin{bmatrix}
 & 4 & 0 & 0 & 0 &\\
& 0 & 1 & 0 & 0 &\\
 & 0 & 0 & 0 & 0 & \\
 & 0& 0& 0 & 4 &
\end{bmatrix},
\end{equation}
where I include $\nu_i$ with a variance of zero such that agents have perfect information on prices.

\begin{figure}[hbtp!]
\centering
\includegraphics[width=\linewidth]{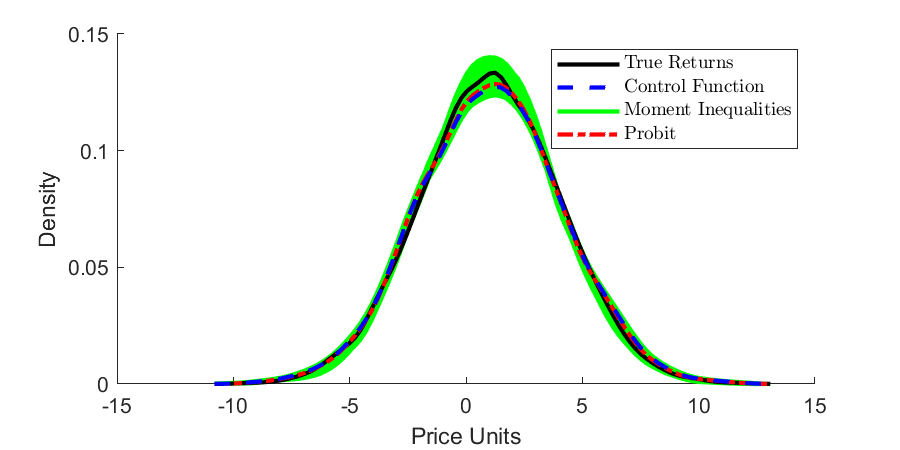}
\begin{minipage}{1\linewidth}
\caption{Perceived Returns Distribution, Known Exogenous Prices}
\label{fig:sim_1}
\smallskip
\footnotesize
\emph{Notes:} Estimated densities of perceived returns given by each method. Densities for each parameter vector in the moment inequalities' 95\% confidence set are shown using $\varphi=[0,1]$ with steps of $1/4$.
\end{minipage}
\end{figure}

Table \ref{tab:sim_1} shows perceived returns estimates for one simulation of this DGP using all three methods. Figure \ref{fig:sim_1} shows the distributions implied by the estimates for each method. Because this DGP is particularly well-behaved, all three methods' estimates are very close to the data-generating parameters.
Additionally, the moment inequalities provide very tight bounds here because the first-stage error has relatively low variance such that making use of $\mathbb{E}[Price_i|Z_i]$ in place of $\widetilde{Price}_i$ introduces little uncertainty into the estimated perceived returns.

\begin{table}[htbp]\centering
\def\sym#1{\ifmmode^{#1}\else\(^{#1}\)\fi}
\caption{Perceived Returns Estimates, Known Exogenous Prices}
\label{tab:sim_1}
\begin{tabular}{l*{1}|c|*{3}{c}}
\hline\hline
                    &            &\multicolumn{1}{c}{(1)}&\multicolumn{1}{c}{(2)}&\multicolumn{1}{c}{(3)}\\
                    &   Target   &\multicolumn{1}{c}{Probit}&\multicolumn{1}{c}{Control Function}&\multicolumn{1}{c}{Moment Inequalities}\\
\hline
Constant            &     1      &     0.986  &      0.997 &  [0.906,    1.066]          \\
                    &            &    (0.033) &     (0.034)&        N/A    \\
$\sigma$            &     2      &     2.071  &          . &         .   \\
                    &            &    (0.040) &         &            \\
$\sigma_\zeta$      &   2.012    &         .  &      2.092 &        .    \\
                    &            &            &     (0.043)&            \\
$\rho$              &     0      &         .  &     -0.051 &        .    \\
                    &            &            &     (0.036)&            \\
$(\sigma_\epsilon, \sigma_\xi)$&   (2,2)    &         .  &          . &   [1.976,    2.236]         \\
                    &            &            &        &     N/A       \\
\hline
Observations        &            &       10000&       10000&       10000\\

\hline\hline
\end{tabular}
\begin{minipage}{1\linewidth}
\smallskip
\footnotesize
\emph{Notes:} Standard errors in parentheses, corrected for the inclusion of estimated regressors following \cite{mt85} in the case of the control function. 
Parameters are in monetary units.
Estimates relate to expressions (\ref{probitdist}), (\ref{CF_dist}), and (\ref{MIdist}), respectively.
The moment inequalities estimate bounds for $\sigma_\epsilon$ under the assumptions in Appendix \ref{appendix:Moment_Inequalities} and $\sigma_\xi$ under those in Appendix \ref{appendix:MI_Proofs}.
All data is generated in Stata using random seed 1234.
\end{minipage}
\end{table}

\subsection{Mean-reverting Misperceptions of Exogenous Prices}
In this simulation, I consider a DGP that corresponds to the setting described in Section \ref{MI_method} in which agents do not precisely forecast prices such that $Price_i \neq \widetilde{Price}_i$. Specifically, price misperceptions move in the opposite direction of prices such that $Cov(Price_i,\nu_i)<0$, as in the case when agents form rational expectations on prices using a strict subset of relevant forecasting variables. This causes agents to tend to believe their price is closer to the average than it actually is.
I generate data according to
\begin{equation}
\begin{bmatrix}
z_i \\ 
u_i \\
\nu_i \\
\epsilon_i
\end{bmatrix}
\sim 
\mathcal{N}(\textbf{0},\Sigma);
\quad \Sigma = 
\begin{bmatrix}
&4 & 0 & 0 & 0&\\
&0 & 7 & -7 & 0&\\
& 0 & -7 & 12 & 0& \\
& 0& 0& 0 & 4&
\end{bmatrix}.
\end{equation}

\begin{figure}[hbtp!]
\centering
\includegraphics[width=\linewidth]{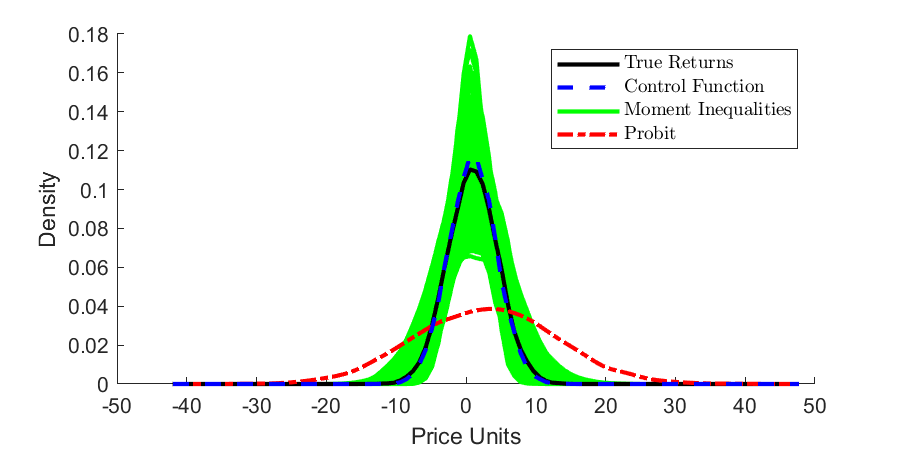}
\begin{minipage}{1\linewidth}
\caption{Perceived Returns Distributions, Mean-Reverting Misperceptions}
\label{fig:sim_2}
\smallskip
\footnotesize
\emph{Notes:} Estimated densities of perceived returns given by each method. Densities for each parameter vector in the moment inequalities' 95\% confidence set are shown using $\varphi=[0,1]$ with steps of $1/4$.
\end{minipage}
\end{figure}

Table \ref{tab:sim_2} shows the estimates for one simulation of this DGP using all three methods. Figure \ref{fig:sim_2} shows the distributions implied by the estimates for each method. The control function and moment inequality estimates are close to the true parameters. 
The control function estimates are significantly more precise than those of the moment inequalities. The probit's estimates are biased upward as expected, given the normality assumptions on the errors \citep{yg85}.

\begin{table}[htbp]\centering
\def\sym#1{\ifmmode^{#1}\else\(^{#1}\)\fi}
\caption{Perceived Returns Estimates, Mean-Reverting Misperceptions}
\label{tab:sim_2}
\begin{tabular}{l*{1}|c|*{3}{c}}
\hline\hline
                    &            &\multicolumn{1}{c}{(1)}&\multicolumn{1}{c}{(2)}&\multicolumn{1}{c}{(3)}\\
                    &   Target   &\multicolumn{1}{c}{Probit}&\multicolumn{1}{c}{Control Function}&\multicolumn{1}{c}{Moment Inequalities}\\
\hline
Constant            &     1      &     2.753  &      0.896 &  [-0.819,    2.611]          \\
                    &            &    (0.160) &     (0.071)&      N/A      \\
$\sigma$            &     4      &     9.527  &          . &        .    \\
                    &            &    (0.373) &         &            \\
$\sigma_\zeta$      &   2.966    &         .  &      2.808 &        .    \\
                    &            &            &     (0.106)&            \\
$\rho$              &     1      &         .  &      1.014 &        .    \\
                    &            &            &     (0.020)&            \\
$(\sigma_\epsilon, \sigma_\xi)$&   (2,3)    &         .  &          . &  [1.119,    4.919]          \\
                    &            &            &         &        N/A    \\
\hline
Observations        &            &       10000&       10000&       10000\\

\hline\hline
\end{tabular}
\begin{minipage}{1\linewidth}
\smallskip
\footnotesize
\emph{Notes:} Standard errors in parentheses, corrected for the inclusion of estimated regressors following \cite{mt85} in the case of the control function. 
Parameters are in monetary units.
Estimates relate to expressions (\ref{probitdist}), (\ref{CF_dist}), and (\ref{MIdist}), respectively.
The moment inequalities estimate bounds for $\sigma_\epsilon$ under the assumptions in Appendix \ref{appendix:Moment_Inequalities} and $\sigma_\xi$ under those in Appendix \ref{appendix:MI_Proofs}.
All data is generated in Stata using random seed 1234.
\end{minipage}
\end{table}

\subsection{Known, Positively Selected Prices}
In this simulation, I consider a DGP in which prices are known, but $u_i$ and $\epsilon_i$ are positively correlated, such as in the case of price discrimination.
This setting is one case of that described in Section \ref{CF_method}.
I generate data according to
\begin{equation}
\begin{bmatrix}
z_i \\ 
u_i \\
\nu_i \\
\epsilon_i
\end{bmatrix}
\sim 
\mathcal{N}(\textbf{0},\Sigma);
\quad \Sigma = 
\begin{bmatrix}
 & 4 & 0 & 0 & 0&\\
& 0 & 7 & 0 & 0&\\
 & 0 & 0 & 0 & 7& \\
 & 0& 0& 7 & 16&
\end{bmatrix}.
\end{equation}

\begin{figure}[hbtp!]
\centering
\includegraphics[width=\linewidth]{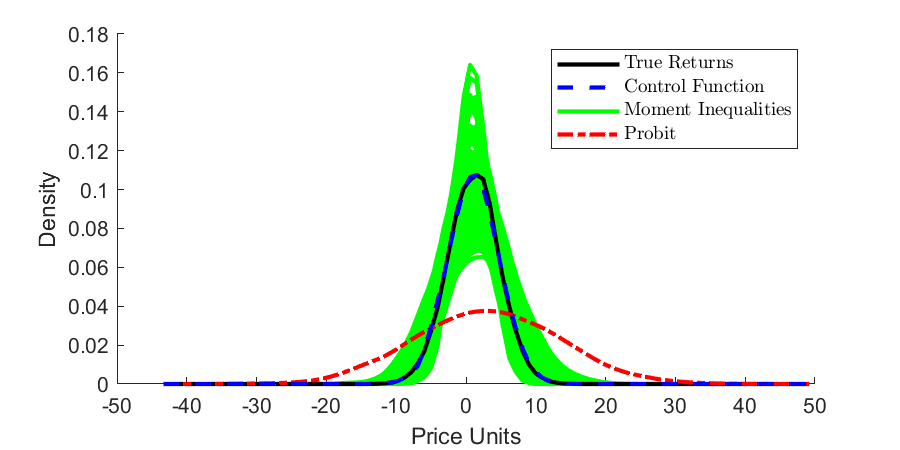}
\begin{minipage}{1\linewidth}
\caption{Perceived Returns Distributions, Positively Selected Prices}
\label{fig:sim_A3}
\smallskip
\footnotesize
\emph{Notes:} Estimated densities of perceived returns given by each method. Densities for each parameter vector in the moment inequalities' 95\% confidence set are shown using $\varphi=[0,1]$ with steps of $1/4$.
\end{minipage}
\end{figure}

Table \ref{tab:sim_A3} shows the estimates for one simulation of this DGP using all three methods. Figure \ref{fig:sim_A3} shows the distributions implied by the estimates for each method. The control function estimates are close to the true parameter values, while the moment inequalities also bound the true parameters. The probit estimates are biased, as expected given the price endogeneity.

In this case $\rho\in[0,1]$, which is relevant for the performance of the moment inequalities, as described in Appendix \ref{appendix:MI_Proofs}). In short, this produces correlation between prices and omitted variables that is functionally equivalent to that of classical measurement error in prices as measures of perceived prices as described in Section \ref{MI_method}. Regarding the value of $\rho$, it is worth noting that the sign is determined by $\mathbb{E}[u_i(-\nu_i+\epsilon_i)]$, such that negative (positive) correlation between $u_i$ and $\nu_i$ will produce an equivalent situation as positive (negative) correlation between $u_i$ and $\epsilon_i$.

\begin{table}[htbp]\centering
\def\sym#1{\ifmmode^{#1}\else\(^{#1}\)\fi}
\caption{Perceived Returns Estimates, Positively Selected Prices}
\label{tab:sim_A3}
\begin{tabular}{l*{1}|c|*{3}{c}}
\hline\hline
                    &            &\multicolumn{1}{c}{(1)}&\multicolumn{1}{c}{(2)}&\multicolumn{1}{c}{(3)}\\
                    &   Target   &\multicolumn{1}{c}{Probit}&\multicolumn{1}{c}{Control Function}&\multicolumn{1}{c}{Moment Inequalities}\\
\hline
Constant            &     1      &     2.932  &      0.989 &  [-0.483,    2.757]          \\
                    &            &    (0.168) &     (0.074)&     N/A       \\
$\sigma$            &     4      &     9.747  &          . &       .     \\
                    &            &    (0.386) &         &            \\
$\sigma_\zeta$      &   3.026    &         .  &      3.080 &      .      \\
                    &            &            &     (0.113)&            \\
$\rho$              &     1      &         .  &      1.007 &      .      \\
                    &            &            &     (0.021)&            \\
$(\sigma_\epsilon, \sigma_\xi)$&   (4,3)    &         .  &          . &  [1.271,    4.889]          \\
                    &            &            &         &     N/A       \\
\hline
Observations        &            &       10000&       10000&       10000\\

\hline\hline
\end{tabular}
\begin{minipage}{1\linewidth}
\smallskip
\footnotesize
\emph{Notes:} Standard errors in parentheses, corrected for the inclusion of estimated regressors following \cite{mt85} in the case of the control function. 
Parameters are in monetary units.
Estimates relate to expressions (\ref{probitdist}), (\ref{CF_dist}), and (\ref{MIdist}), respectively.
The moment inequalities estimate bounds for $\sigma_\epsilon$ under the assumptions in Appendix \ref{appendix:Moment_Inequalities} and $\sigma_\xi$ under those in Appendix \ref{appendix:MI_Proofs}.
All data is generated in Stata using random seed 1234.
\end{minipage}
\end{table}

\subsection{Known, Negatively Selected Prices}

In this simulation, I consider a DGP in which $u_i$ and $\epsilon_i$ are negatively correlated. This setting is a case of the one described in Section \ref{CF_method}.
I generate data according to
\begin{equation}
\begin{bmatrix}
z_i \\ 
u_i \\
\nu_i \\
\epsilon_i
\end{bmatrix}
\sim 
\mathcal{N}(\textbf{0},\Sigma);
\quad \Sigma = 
\begin{bmatrix}
 & 4 & 0 & 0 & 0&\\
& 0 & 20 & 0 & -10&\\
 & 0 & 0 & 0 & 0& \\
 & 0& -10& 0 & 9&
\end{bmatrix}.
\end{equation}
where I include $\nu_i$ with a variance of zero to emphasize that there are no price misperceptions in this case.

\begin{figure}[hbtp!]
\centering
\includegraphics[width=\linewidth]{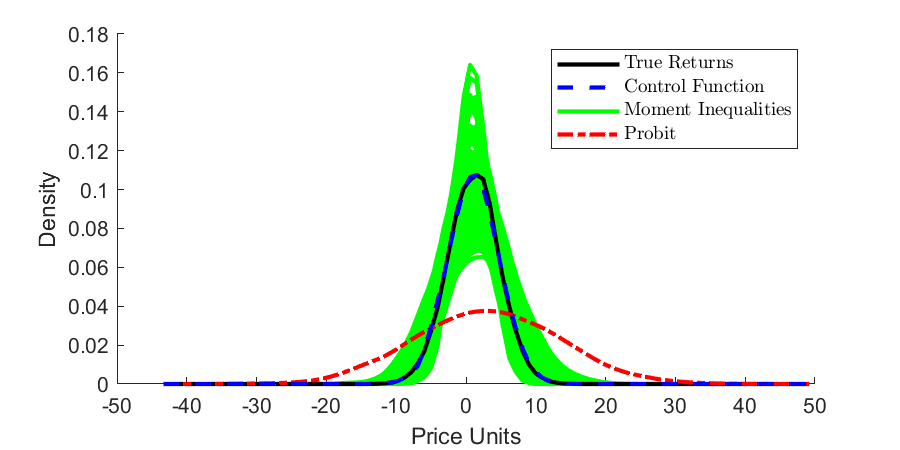}
\begin{minipage}{1\linewidth}
\caption{Perceived Returns Distributions, Negatively Selected Prices}
\label{fig:sim_A4}
\smallskip
\footnotesize
\emph{Notes:} Estimated densities of perceived returns given by each method. Densities for each parameter vector in the moment inequalities' 95\% confidence set are shown using $\varphi=[0,1]$ with steps of $1/4$.
\end{minipage}
\end{figure}

Table \ref{tab:sim_A4} shows the estimates for one simulation of this DGP using all three methods. Figure \ref{fig:sim_A4} shows the distributions implied by the estimates for each method. The control function method estimates are close to the true parameter values, while the other methods perform poorly. In the case of the probit, there is nothing to address inequality, while the moment inequalities address positive correlation between prices and the composite idiosyncratic preference term $\-nu_i+\epsilon_i$, but not negative correlation.

\begin{table}[htbp]\centering
\def\sym#1{\ifmmode^{#1}\else\(^{#1}\)\fi}
\caption{Perceived Returns Estimates, Negatively Selected Prices}
\label{tab:sim_A4}
\begin{tabular}{l*{1}|c|*{3}{c}}
\hline\hline
                    &            &\multicolumn{1}{c}{(1)}&\multicolumn{1}{c}{(2)}&\multicolumn{1}{c}{(3)}\\
                    &   Target   &\multicolumn{1}{c}{Probit}&\multicolumn{1}{c}{Control Function}&\multicolumn{1}{c}{Moment Inequalities}\\
\hline
Constant            &     1      &     0.739  &      1.100 &  [-1.753,    2.654]          \\
                    &            &    (0.034) &     (0.088)&     N/A       \\
$\sigma$            &     3      &     1.554  &          . &     .       \\
                    &            &    (0.032) &         &            \\
$\sigma_\zeta$      &   2.019    &         .  &      2.011 &     .       \\
                    &            &            &     (0.099)&            \\
$\rho$              &    -.5     &         .  &     -0.497 &      .      \\
                    &            &            &     (0.059)&            \\
$(\sigma_\epsilon, \sigma_\xi)$&   (3,2)    &         .  &          . &  [0.131,   0.922]          \\
                    &            &            &        &   N/A         \\
\hline
Observations        &            &       10000&       10000&       10000\\

\hline\hline
\end{tabular}
\begin{minipage}{1\linewidth}
\smallskip
\footnotesize
\emph{Notes:} Standard errors in parentheses, corrected for the inclusion of estimated regressors following \cite{mt85} in the case of the control function. 
Parameters are in monetary units.
Estimates relate to expressions (\ref{probitdist}), (\ref{CF_dist}), and (\ref{MIdist}), respectively.
The moment inequalities estimate bounds for $\sigma_\epsilon$ under the assumptions in Appendix \ref{appendix:Moment_Inequalities} and $\sigma_\xi$ under those in Appendix \ref{appendix:MI_Proofs}.
All data is generated in Stata using random seed 1234.
\end{minipage}
\end{table}

\subsection{Mean-reverting Misperceptions of Positively Selected Prices}\label{sim_EI}
In this simulation, I consider a DGP in which $u_i$ is positively correlated with $\epsilon_i$ and negative correlated with $\nu_i$. This case would occur in a setting in which there is price discrimination on unobserved components of preferences, and agents are only aware of a subset of price determinants and form rational expectations based on known price determinants.
I generate data according to
\begin{equation}
\begin{bmatrix}
z_i \\ 
u_i \\
\nu_i \\
\epsilon_i
\end{bmatrix}
\sim 
\mathcal{N}(\textbf{0},\Sigma);
\quad \Sigma = 
\begin{bmatrix}
 & 4 & 0 & 0 & 0&\\
& 0 & 7 & -3 & 4&\\
 & 0 & -3 & 12 & 0& \\
 & 0& 4& 0 & 4&
\end{bmatrix}.
\end{equation}
This setting corresponds to the one described in Section \ref{CF_method}. This setting is likely the most realistic, insofar as mean-reverting price misperceptions and positive selection on prices are likely. In this case, Section \ref{CF_method} and Appendix \ref{appendix:MI_Proofs} suggest that the control function method and the moment inequality method will consistently estimate perceived returns, but the probit will not.

\begin{figure}[hbtp!]
\centering
\includegraphics[width=\linewidth]{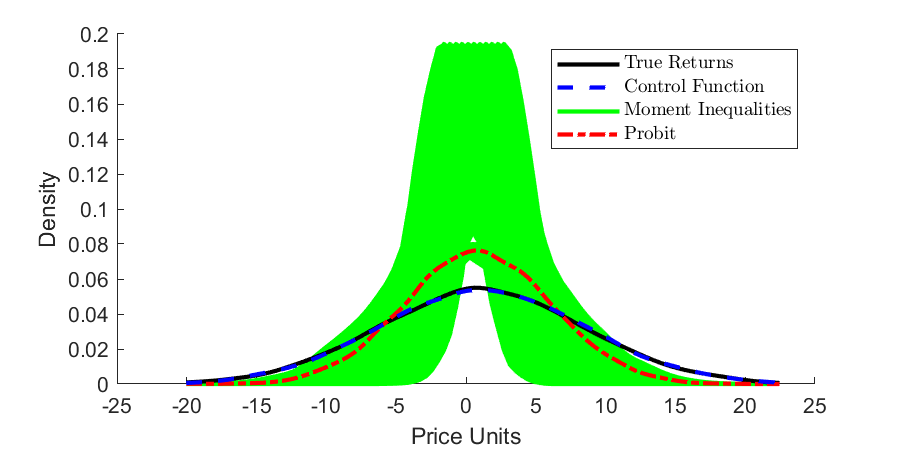}
\begin{minipage}{1\linewidth}
\caption{Perceived Returns Distributions, Positively Selected Partially Known Prices}
\label{fig:sim_A5}
\smallskip
\footnotesize
\emph{Notes:} Estimated densities of perceived returns given by each method. Densities for each parameter vector in the moment inequalities' 95\% confidence set are shown using $\varphi=[0,1]$ with steps of $1/4$.
\end{minipage}
\end{figure}

Table \ref{tab:sim_A5} shows the estimates for one simulation of this DGP using all three methods. Figure \ref{fig:sim_A5} shows the distributions implied by the estimates for each method. The control function method estimates are close to the true parameter values, while the moment inequalities bound the true values. The probit estimates are biased away from zero, because variation in prices predicts relatively modest changes in investment, as not all price variation is known to agents and because price variation is accompanied by higher idiosyncratic preferences for investment.

\begin{table}[htbp]\centering
\def\sym#1{\ifmmode^{#1}\else\(^{#1}\)\fi}
\caption{Perceived Returns Estimates, Positively Selected Partially Known Prices}
\label{tab:sim_A5}
\begin{tabular}{l*{1}|c|*{3}{c}}
\hline\hline
                    &            &\multicolumn{1}{c}{(1)}&\multicolumn{1}{c}{(2)}&\multicolumn{1}{c}{(3)}\\
                    &   Target   &\multicolumn{1}{c}{Probit}&\multicolumn{1}{c}{Control Function}&\multicolumn{1}{c}{Moment Inequalities}\\
\hline
Constant            &     1      &     3.076  &      0.978 &  [-0.785,    2.740]          \\
                    &            &    (0.173) &     (0.073)&      N/A      \\
$\sigma$            &     4      &     9.851  &          . &        .    \\
                    &            &    (0.399) &         (.)&            \\
$\sigma_\zeta$      &   2.982    &         .  &      2.824 &        .    \\
                    &            &            &     (0.107)&            \\
$\rho$              &     1      &         .  &      1.027 &        .    \\
                    &            &            &     (0.020)&            \\
$(\sigma_\epsilon, \sigma_\xi)$&   (2,3)    &         .  &          . &  [1.107,    4.970]          \\
                    &            &            &         (.)&      N/A      \\
\hline
Observations        &            &       10000&       10000&       10000\\

\hline\hline
\end{tabular}
\begin{minipage}{1\linewidth}
\smallskip
\footnotesize
\emph{Notes:} Standard errors in parentheses, corrected for the inclusion of estimated regressors following \cite{mt85} in the case of the control function. 
Parameters are in monetary units.
Estimates relate to expressions (\ref{probitdist}), (\ref{CF_dist}), and (\ref{MIdist}), respectively.
The moment inequalities estimate bounds for $\sigma_\epsilon$ under the assumptions in Appendix \ref{appendix:Moment_Inequalities} and $\sigma_\xi$ under those in Appendix \ref{appendix:MI_Proofs}.
All data is generated in Stata using random seed 1234.
\end{minipage}
\end{table}

\subsection{Computational Comparison with Controls}

Next, I present two simulations which include additional explanatory variables. This exercise is intended to provide a computational comparison of the control function method and the moment inequality method, so they include the computation time taken to complete each procedure. These simulations use the DGP described in Section \ref{sim_EI} with the addition of the variables $x_1$ in the first simulation, and $x_1$ and $x_2$ in the second. I set $x_1 \sim \mathcal{N}(0,4)$ and $x_2 \sim \mathcal{N}(0,4)$ with coefficients of zero. The results are shown in Table \ref{tab:sim_A6} and Table \ref{tab:sim_A7}, respectively, where graphs of implied perceived returns are omitted because they are visually indistinguishable from Figure \ref{fig:sim_A5} (given the zero coefficients on the new variables). All simulations are performed on a Linux server with two Intel Xeon X5550 CPUs and 48GB of RAM. Note that the run times in seconds for the moment inequalities are orders of magnitude higher than the other methods for both simulations, and that this difference is increasing in the number of variables.

\begin{table}[htbp]\centering
\def\sym#1{\ifmmode^{#1}\else\(^{#1}\)\fi}
\caption{Perceived Returns Estimates, 1 Control}
\label{tab:sim_A6}
\begin{tabular}{l*{4}{c}}
\hline\hline
                    &            &\multicolumn{1}{c}{(1)}&\multicolumn{1}{c}{(2)}&\multicolumn{1}{c}{(3)}\\
                    &   Target   &\multicolumn{1}{c}{Probit}&\multicolumn{1}{c}{Control Function}&\multicolumn{1}{c}{Moment Inequalities}\\
\hline
Constant            &     1      &     3.075  &      0.977 &  [-6.956, 8.029]          \\
                    &            &    (0.173) &     (0.073)&  N/A         \\
$x_1        $         &     0      &    -0.008  &     -0.008 &  [-0.795, 0.778]          \\
                    &            &    (0.063) &     (0.033)&  N/A          \\
$\sigma$            &     4      &     9.851  &          . &      .      \\
                    &            &    (0.399) &         &            \\
$\sigma_\zeta$      &   2.982    &         .  &      2.823 &        .    \\
                    &            &            &     (0.107)&            \\
$\rho$              &     1      &         .  &      1.027 &        .    \\
                    &            &            &     (0.020)&            \\
$(\sigma_\epsilon, \sigma_\xi)$&   (2,3)    &         .  &          . &   [1.106, 4.970]         \\
                    &            &            &          &   N/A         \\
\hline
Observations        &            &       10000&       10000&       10000\\
Computation Time       &            &     0      &       2 &       1017\\
\hline\hline
\end{tabular}
\begin{minipage}{1\linewidth}
\smallskip
\footnotesize
\emph{Notes:} Standard errors in parentheses, corrected for the inclusion of estimated regressors following \cite{mt85} in the case of the control function. 
Parameters are in monetary units.
Estimates relate to expressions (\ref{probitdist}), (\ref{CF_dist}), and (\ref{MIdist}), respectively.
The moment inequalities estimate bounds for $\sigma_\epsilon$ under the assumptions in Appendix \ref{appendix:Moment_Inequalities} and $\sigma_\xi$ under those in Appendix \ref{appendix:MI_Proofs}.
All data is generated in Stata using random seed 1234.
Computation time is rounded to the nearest whole second.
\end{minipage}
\end{table}

\begin{table}[htbp]\centering
\def\sym#1{\ifmmode^{#1}\else\(^{#1}\)\fi}
\caption{Perceived Returns Estimates, 2 Controls}
\label{tab:sim_A7}
\begin{tabular}{l*{4}{c}}
\hline\hline
                    &            &\multicolumn{1}{c}{(1)}&\multicolumn{1}{c}{(2)}&\multicolumn{1}{c}{(3)}\\
                    &   Target   &\multicolumn{1}{c}{Probit}&\multicolumn{1}{c}{Control Function}&\multicolumn{1}{c}{Moment Inequalities}\\
\hline
Constant            &     1      &     3.076  &      0.977 &  [-14.890, 16.550]          \\
                    &            &    (0.173) &     (0.073)&     N/A       \\
$x_1    $             &     0      &    -0.009  &     -0.008 &  [-1.189, 1.303]          \\
                    &            &    (0.064) &     (0.033)&            \\
$x_2    $             &     0      &     0.030  &      0.002 &  [-1.327, 1.198]          \\
                    &            &    (0.064) &     (0.033)&    N/A        \\
$\sigma$            &     4      &     9.852  &          . &      .      \\
                    &            &    (0.399) &         &            \\
$\sigma_\zeta$      &   2.982    &         .  &      2.823 &       .     \\
                    &            &            &     (0.107)&            \\
$\rho$              &     1      &         .  &      1.027 &       .     \\
                    &            &            &     (0.020)&            \\
$(\sigma_\epsilon, \sigma_\xi)$&   (2,3)    &         .  &          . &   [0.677, 4.970]         \\
                    &            &            &         &       N/A     \\
\hline
Observations        &            &       10000&       10000&       10000\\

Computation Time     &            &     1      &       2 &       21085
\\
\hline\hline
\end{tabular}
\begin{minipage}{1\linewidth}
\smallskip
\footnotesize
\emph{Notes:} Standard errors in parentheses, corrected for the inclusion of estimated regressors following \cite{mt85} in the case of the control function. 
Parameters are in monetary units.
Estimates relate to expressions (\ref{probitdist}), (\ref{CF_dist}), and (\ref{MIdist}), respectively.
The moment inequalities estimate bounds for $\sigma_\epsilon$ under the assumptions in Appendix \ref{appendix:Moment_Inequalities} and $\sigma_\xi$ under those in Appendix \ref{appendix:MI_Proofs}.
All data is generated in Stata using random seed 1234.
Computation time is rounded to the nearest whole second.
\end{minipage}
\end{table}

\end{appendices}

\end{document}